\documentclass[letterpaper, 11pt]{article}
\usepackage{amsmath}
\usepackage{amsfonts}
\usepackage{amsthm}
\usepackage{amssymb}
\usepackage{mathrsfs}
\usepackage[hang]{subfigure}
\usepackage{graphicx,epsfig,fancyheadings,wasysym,psfrag}

\usepackage{times}

\usepackage{natbib}
\DeclareMathOperator{\var}{var}
\DeclareMathOperator{\cov}{cov}

\newcommand{\al}{\alpha}
\newcommand{\be}{\beta}
\newcommand{\de}{\delta}
\newcommand{\e}{\epsilon}

\newcommand{\la}{\lambda}

\newcommand{\cG}{{\mathcal G}}

\newcommand{\cN}{{\mathcal N}}

\newcommand{\cO}{{\mathcal O}}
\newcommand{\cP}{{\mathcal P}}

\newcommand{\conn}{\leftrightsquigarrow}
\newcommand{\notconn}{{\,\,\leftrightsquigarrow\!\!\!\!\!\!\!\!/\;\,\,\,}}

\newcommand{\aas}{a.~a.~s.}

\begin{document}

\title{Percolation on fitness landscapes: effects of correlation, phenotype, and incompatibilities}
\author{Janko Gravner$^*$, Damien Pitman$^*$, and Sergey Gavrilets$^{\dag\ddag}$\\
$^*$Department of Mathematics, University of California, Davis, CA 95616,\\
$^{\dag}$Departments of Ecology and Evolutionary Biology
and Mathematics, \\
University of Tennessee, Knoxville, TN 37996, USA.\\ 
$^\ddag$corresponding author.
Phone: 865-974-8136,\
fax: 865-974-3067,\\
email: gavrila@tiem.utk.edu}

\maketitle

\newpage

{\bf Abstract}\quad
We study how correlations in the random fitness assignment may affect the structure
of fitness landscapes. We consider three classes of fitness models. The 
first is a continuous phenotype space in which individuals are characterized 
by a large number of continuously varying traits such as size, weight, color, or 
concentrations of gene products which directly affect fitness.
The second is a simple model that explicitly describes genotype-to-phenotype 
and phenotype-to-fitness maps allowing for neutrality at both phenotype and fitness
levels and resulting in a fitness landscape with tunable correlation length.
The third is a class of models in which particular combinations of alleles or 
values of phenotypic characters are ``incompatible'' in the sense that the
resulting genotypes or phenotypes have reduced (or zero) fitness. 
This class of models
can be viewed as a generalization of the canonical Bateson-Dobzhansky-Muller
model of speciation.
We also demonstrate that the discrete $NK$ model shares some signature properties of models 
with high correlations.
Throughout the paper, our focus is on the percolation threshold, on the number, size and 
structure of connected clusters, and on the number of viable genotypes. \\


\section{Introduction}

The notion of fitness landscapes, introduced by a theoretical evolutionary biologist Sewall
Wright in \citeyear{wri32} (see also \citealt{kau93,gav04}), has proved extremely useful both in
biology and well outside of it. In the standard interpretation, a fitness landscape is a relationship
between a set of genes (or a set of quantitative characters) and a measure of fitness
(e.g. viability, fertility, or mating success). In Wright's original formulation the set of 
genes (or quantitative characters) is the property of an individual. However, the notion of 
fitness landscapes can be generalized to the level of a mating pair, or even a population of
individuals \citep{gav04}.

To date, most empirical information on fitness landscapes in biological applications has come from studies 
of RNA (e.g., \citealt{sch95,huy96b,fon98b}), 
proteins (e.g., \citealt{lip91,mar96,ros97}), 
viruses (e.g.,  \citealt{bur99,bur04}), 
bacteria (e.g., \citealt{ele03,woo06}),
and artificial life (e.g.,  \citealt{len99,wil01c}).
The three paradigmatic landscapes --- rugged, single-peak, 
and flat --- emphasizing particular
features of fitness landscapes have been the focus of most of the earlier theoretical work 
(reviewed in \citealt{kau93,gav04}). These landscapes have found numerous applications with regards to the dynamics
of adaptation (e.g., \citealt{kau87,kau93,orr06a,orr06b}) 
and neutral molecular evolution (e.g., \citealt{der91}).

More recently, it was realized that the dimensionality of most biologically interesting
fitness landscapes is enormous and that this huge dimensionality brings some new properties 
which one does not observe in low-dimensional landscapes (e.g. in two- or three-dimensional 
geographic landscapes). In particular, multidimensional landscapes are generically characterized
by the existence of neutral and nearly neutral networks (also referred to as holey fitness
landscapes) that extend throughout the landscapes
and that can dramatically affect the evolutionary dynamics of the populations
\citep{gav97,gav97b,rei97b,gav04,rei01a,rei01b,rei02}.

An important property of fitness landscapes is their correlation pattern. A common measure 
for the strength of dependence 
is the {\it correlation function\/} $\rho$ measuring the correlation of 
fitnesses  of pairs of individual at a distance (e.g., Hamming) $d$ from each other in the 
genotype (or phenotype) space:
	\begin{equation} \label{rho}
		\rho(d)=\frac{\cov[w(.),w(.)]_d}{\var(w)}
	\end{equation}
\citep{eig89}. Here, the term in the numerator is the covariance of fitnesses 
of two individuals conditioned on them being at distance $d$, and
$\var(w)$ is the variance in fitness over the whole fitness landscape.
For uncorrelated landscapes, $\rho(d)=0$ for $d > 0$. In contrast,
for highly correlated landscapes, $\rho(d)$ decreases with $d$ very slowly.

The aim of this paper is to extend our previous work \citep{gav97b} in a number of directions
paying special attention to the question of how correlations in the
random fitness assignment may affect the structure of genotype and phenotype spaces. 
For the resulting random fitness landscapes, we shed some 
light on issues such as the number of viable genotypes, 
number of connected clusters of viable genotypes and 
their size distribution, existence thresholds, and 
number of possible fitnesses. 

To this end, we introduce a variety of models, 
which could be divided into two essentially different 
classes: those with local correlations, and 
those with global correlations. As we will see, techniques 
used to analyze these models, and answers we obtain, differ 
significantly. We use a mixture of analytical and computational techniques; 
it is perhaps necessary to point out that these models
are very far from trivial, and one is quickly led to 
outstanding open problems in probability theory and computer science. 

We start (in Section 2) by briefly reviewing some results from \cite{gav97b}.
In Section 3 we generalize these results for the case of a continuous
phenotype space when individuals are characterized by a large number 
of continuously varying traits such as size, weight, color, or the 
concentrations of some gene products. The latter interpretation
of the phenotype space may be particularly relevant given the rise of
proteomics and the growing interest in gene regulatory networks.

The main idea behind our local correlations model studies in Section 4
is fitness assignment {\it conformity\/}. Namely, one randomly divides 
the genotype space into components which are forced to have 
the same phenotype; then, each different phenotype is independently assigned a random fitness. 
This leads to a simple two-parameter 
model, in which one parameter determines the density of viable genotypes,  
and the other the correlations between them. 
We argue that the probability of existence of a giant cluster (which swallows a positive 
proportion of all viable genotypes) is a non-monotone function of the correlation 
parameter and identify the critical surface at which this probability jumps 
almost from 0 to 1. In Section 4 we also investigate the effects of 
interaction between conformity structure and fitness assignment. 

Section 5 introduces our basic global correlation 
model, one in which genotypes are eliminated due to random pairwise
{\it incompatibilities\/} between alleles. This is 
equivalent to a random version of {\tt SAT} problem, 
which is the canonical constraint satisfaction problem in computer 
science. In general, a {\tt SAT} problem involves a set of Boolean variables
and their negations that are strung together with {\tt OR} symbols into
{\it clauses\/}.  The {\it clauses\/} are joined by {\tt AND} symbols
into a {\it formula\/}. A {\tt SAT} problem asks one to decide, whether
the variables can be assigned values that will make the formula true. 
An important special case, $K$-{\tt SAT}, has the length of each clause fixed at $K$. 
Arguably, {\tt SAT} is the most important class of problems in complexity theory.
In fact, the general {\tt SAT} was the first known
NP-complete problem and was established as such by S. Cook in 1971 (\citealt{Coo}). 
Even considerable simplifications, such as the {\tt $3$-SAT} (see Section 5.4), remain NP-complete,
although {\tt $2$-SAT} (see Section 5.1) can be solved efficiently by a simple algorithm.
See e.g. \cite{KV} for a comprehensive presentation of the theory. Difficulties
in analyzing random  {\tt SAT} problems, in which formulas are chosen at random,
in many ways mirror their complexity classes, but even random {\tt $2$-SAT}
presents significant challenges \citep{dlV, BKL2}. In our present interpretation, the main reason
for these difficulties is that correlations are so high that the expected number
of viable genotypes
may be exponentially large, while at the same time the probability 
that even one viable genotype exists is very low. In Section 5, we further 
illuminate this issue by showing that connected viable clusters 
must contain fairly large sub-cubes, and that the number of such clusters 
is, in a proper interpretation, finite. The relevance to both types of 
models for discrete and continuous   
phenotype spaces is also discussed, with particular emphasis on the 
existence of viable phenotypes in the presence of incompatibilities. 
Section 5 also contains a brief review 
of the existing theory on higher order incompatibilities.

In Section 6 we demonstrate how the discrete 
$NK$ model shares some signature properties of models 
with high correlations. In Section 7 we summarize our results 
and discuss their biological relevance. 
The proofs of our major results are relegated to Appendices A--E.

\section{The basic case: binary hypercube and independent binary fitness} 

We begin with a brief review of the basic setup, from \cite{gav97b} 
and \cite{gav04}. The {\it binary hypercube\/}   
consists of all $n$--long arrays of bits, or {\it alleles\/}, that is 
$\cG=\{0, 1\}^n$. This is our {\it genotype space\/}. 
Genotypes are linked by edges induced by bit-flips, i.e., {\it mutations\/} at a single locus, 
for example, for $n=4$, a sequence of mutations might look like \[ 0000\leftrightarrow 1000\leftrightarrow 1001\leftrightarrow 1101\leftrightarrow 1100. 
\]
The (Hamming) {\it distance\/} $d(x,y)$ between $x\in \cG$ and $y\in \cG$ is the 
number of coordinates in which $x$ and $y$ differ or, equivalently, 
the least number of mutations which connect $x$ and $y$. 

The {\it fitness\/} of each genotype $x$ is denoted by $w(x)$.
We will describe several ways to prescribe the fitness $w$ at random, according 
to some probability measure $P$ on the $2^{2^n}$ possible assignments. Then we say that 
an event $A_n$ happens {\it asymptotically almost surely\/} (\aas) 
if $P(A_n)\to 1$ as $n\to\infty$. Typically, $A_n$ will capture 
some important property of (random) clusters of genotypes.  

We commonly assume that $w(x)\in \{0,1\}$ so that $x$ is either viable 
($w(x)=1$) or inviable ($w(x)=0$).
As a natural starting point, \cite{gav97b} considered  uncorrelated landscapes, 
in which $w(x)$ is chosen to be 1  with probability $p_v$, for each $x$ independently of 
others. We assume 
this setup for the rest of this section and note that this 
is a well-studied problem in mathematical literature, 
although it presents considerable technical difficulties and 
some issues are still not completely resolved. 

Given a particular fitness assignment, viable genotypes form 
a subset of $\cG$, which is divided into 
connected {\it components\/} or {\it clusters\/}. 
For example, with $n=4$, if $0000$ is viable, but its 4 neighbors
$1000$, $0100$, $0010$, and $0001$ are not, then it is isolated in its own 
cluster.

Perhaps the most basic result determines the {\it connectivity 
threshold\/} \citep{Tom}: when $p_v>1/2$, the set of all viable genotypes is connected a.~a.~s. 
By contrast, when $p_v<1/2$,  the set of viable genotypes is {\it not\/} connected 
{\aas } This is easily understood, as the connectedness is closely linked to 
isolated genotypes, whose expected number is $2^np_v(1-p_v)^n$. This expectation 
makes a transition from exponentially large to exponentially small at $p_v=1/2$. 
The events $\{x$ is isolated$\}$, $x\in \cG$, are only weakly 
correlated, which implies that when $p_v<1/2$ there are exponentially 
many isolated genotypes with high probability, while when $p_v>1/2$, 
a separate argument shows that the event that the set of viable genotypes contains no isolated vertex 
but is not connected becomes very unlikely for large $n$.
This is perhaps the clearest instance of the 
{\it local method\/}: a local property (no isolated genotypes) 
is \aas~equivalent to a global one (connectivity). 
 
Connectivity is clearly too much to ask for, as $p_v$ above $1/2$ is 
not biologically realistic. Instead, one should look for a weaker 
property which has a chance of occurring at small $p_v$. Such a 
property is {\it percolation\/}, a.~k.~a.~existence of the {\it giant component\/}. 
For this, we scale $p_v=\la_v/n$, for a constant $\la_v$. 
When $\la_v>1$, the set of viable genotypes percolates, that is, it a.~a.~s.~contains a 
component of at least $c\cdot n^{-1} 2^n$ genotypes, with all other 
components of at most polynomial (in $n$) size.
When $\la_v<1$,
the largest component is a.~a.~s.~of size $Cn$. Here and below, $c$ and $C$ are
some constants. These are results from \cite{BKL2}. 

The local method that correctly identifies the percolation threshold 
is a little 
more sophisticated than the one for the connectivity threshold, and
uses branching processes with Poisson offspring distribution --- hence we introduce notation   
Poisson($\la$) for a Poisson distribution with mean $\la$. 
Viewed from, say, genotype $0\dots0$, the binary hypercube locally approximates a tree with 
uniform degree $n$. Thus viable genotypes approximate  
a branching process 
in which every node has the number of successors distributed binomially 
with parameters $n-1$ and $p$, hence this random number has mean about $\la_v$ and 
is approximately Poisson($\la_v$). 
When $\la_v>1$, such a branching process survives forever with probability 
$1-\delta>0$, where $\delta=\delta(\la_v)$, and $\delta(\la)$ is given by the 
implicit equation
\begin{equation}\label{delta}
\delta=e^{\la(\delta-1)}.
\end{equation}
(e.g., \citealt{AN}).
Large trees of viable genotypes created by the 
branching processes which emanate from viable genotypes 
merge into a very large (``giant'') connected set. 
On the other hand, when $\la_v<1$ the branching process dies out with probability 1. 

The condition $\la_v>1$ for the existence of the giant component can be loosely
rewritten as
	\begin{equation} \label{basic}
		p_v > \frac{1}{n}.
	\end{equation}
This shows that the larger the dimensionality $n$ of the genotype space, the smaller
values of the probability of being viable $p_v$ will result in the existence of
the giant component. See \cite{gav97b,gav97,gav04,ski04,pig06}  for discussions of biological 
significance and implications of this important result. 

\section{Percolation in a continuous phenotype space}

In this section we will assume that individuals are characterized by $n$ continuous 
traits (such as size, weight, color, or concentrations of particular gene products). 
To be precise, we let $\cP =[0,1]^n$ be the {\em phenotype space}. 

We begin with the extension of the notion of independent viability. 
The most straightforward analogue of the discrete genotype space considered in the 
previous section involves Poisson point location 
in $\cal{P}$, obtained by generating a Poisson($\lambda$) random variable $N$, and then
choosing points  $x_1,\dots,x_N\in \cP$ uniformly at random. 
These will be interpreted as {\it peaks\/} 
of equal height in the fitness landscape. 
Another parameter is a small $r>0$, which can be interpreted as measuring
how harsh the environment is: any phenotype within $r$ 
of one of the peaks is declared viable and any phenotype not within $r$ of one of the peaks
is declared inviable. For simplicity, we will assume ``within 
$r$'' to mean that ``every coordinate differs by at most $r$,''
i.e., distance is measured in the ($n$-dimensional) $\ell^\infty$ norm $||\cdot||_\infty$. 
Note that this makes the set of viable genotypes correlated, albeit 
the range of correlations is limited to $2r$.

Our most basic question is whether a positive proportion of 
viable phenotypes is connected together into a giant cluster.   
Note that the probability $p_v$ that a random point in $\cP$ is viable 
is equal to the probability that there is a ``peak'' within $r$ from this
point. Therefore,
$$
p_v=1-\exp\left[-\lambda (2r)^n\right]\approx \lambda (2r)^n. 
$$
This is also the expected combined volume of viable phenotypes. 
  
We will consider peaks
$x_i$ and $x_j$ to be {\it neighbors\/} if they share a viable phenotype, 
that is, if their $r$-neighborhoods overlap, or 
equivalently, if $||x_i-x_j||_\infty<2r$.  
Two viable phenotypes $y_1$ and $y_2$ are {\it connected\/} if they are, 
respectively, within $r$ of peaks $x_1$ and $x_2$, and $x_1$ and $x_2$ are 
connected to each other via a chain of neighboring peaks.   

By the standard branching process comparison, 
the necessary condition for the existence of a giant cluster is that a ``peak'' $x$ is connected 
to more than one other ``peak'' on the average.
All peaks within $2r$ of the focal peak are connected to the latter.
Therefore, if $\mu$ is the expected number of peaks connected to $x$,
then 
$$
\mu= \lambda \cdot (4r)^n,
$$
and $\mu>1$ is necessary for percolation. 
As demonstrated by \cite{Pen} (for a different choice of 
the norm, but the proof is the same), 
this condition becomes sufficient when $n$ is large. 
Note that the expected number $\lambda$ of peaks can be written as $\mu\cdot (4r)^{-n}$.

If $\mu>1$ and fixed, then \aas~a positive proportion of 
all peaks (that is, $cN$ peaks, where $c=c(\mu)>0$) are connected
in one ``giant'' component, while the remaining connected components are all of size $\cO(\log N)$. 
On the other hand, if $\mu<1$, all components are \aas~of size $\cO(\log N)$.

The condition $\mu>1$ for the existence of the giant component of viable phenotypes can be
loosely rewritten as
	\begin{equation} \label{cont}
		p_v > \frac{1}{2^n}.
	\end{equation}
This shows that viable phenotypes are likely to form a large connected cluster even when  
one is {\it very\/} unlikely to hit one of them at random, if 
$n$ is even moderately large. The same conclusion and the same threshold are valid
if instead of $n$-cubes we use $n$-spheres of a constant radius. 

The percolation threshold in the continuous phenotype space given by inequality~(\ref{cont}) 
is much smaller than that in the discrete genotype space which is given by inequality~(\ref{basic}).
An intuitive reason for this is that continuous space offers a viable point a much greater opportunity
to be connected to a large cluster. Indeed, in the discrete genotype space there are $n$
neighbors per each genotype. In contrast, in the continuous phenotype space, the ratio 
of the volume of the space where neigboring peaks can be located (which has radius $2r$) 
to the volume of the focal $n$-cube (which has radius $r$) is $2^n$.

\section{Percolation in a correlated landscape with phenotypic neutrality}

The standard paradigm in biology is that the relationship between genotype and fitness
is mediated by phenotype (i.e., observable characteristics of individuals). Both the
genotype-to-phenotype and phenotype-to-fitness maps are typically not one-to-one.
Here, we formulate a simple model capturing these properties which also results in a
correlated fitness landscape. 
Below we will call mutations that do not change phenotype {\em conformist}. These mutations
represent a subset of {\em neutral} mutations that do not change fitness.

We propose the following two-step model. To begin the {\it first step\/}, 
we make each  {\it pair\/} of genotypes $x$ and $y$ in a binary hypercube  $\cG$ independently 
{\it conformist\/} with probability $p_{d(x,y)}$ where $d(x,y)$ 
is the Hamming distance between $x$ and $y$. We then declare 
$x$ and $y$ to belong to the same {\it conformist cluster\/} if they are linked 
by a chain of conformist pairs. This version of long-range percolation model (cf., \citealt{Ber,Bis})
divides the set of genotypes $\cG$ into conformist clusters.  
We postulate that all genotypes in the same conformist 
cluster have the same phenotype. Therefore, genetic changes represented by 
a change from one member of a conformist cluster to another (i.e., single or 
multiple mutations) are phenotypically  neutral.

In the {\it second step\/}, we make each conformist cluster independently viable with 
probability $p_v=\la_v/n$. This generates a random set of viable genotypes, 
and we aim to investigate  when this set has a large connected component. 

For example, the ``genotype'' can be a linear RNA sequence.
This sequence folds into a 2-dimensional molecule which has a particular structure 
(or ``shape''), and corresponds to our ``phenotype.'' Finally, the molecule
itself has a particular function, e.g., to bind to a specific part of the cell or
to another molecule. A measure of how well this can be accomplished is represented by
our ``fitness.''

The distribution of conformist clusters depends on the probabilities
$p_1, p_2, p_3, \dots $ which determine how the conformity probability 
varies with distance. 
Here we will study the case when $p_1=p_e>0,p_2=p_3=...=0$ \citep{Hag}. 
It is then very convenient for the mathematical analysis that a pair $x$ and 
$y$ can be conformist only when they are linked by an edge --- therefore 
we can talk about {\it conformist edges\/} or equivalently {\it conformist mutations\/}.  
(Note however that it is possible that nearest neighbors $x$ and $y$ are in the 
same conformist cluster even if the edge between them is non-conformist.) 

Figure 1 illustrates our 2-step procedure on a four-dimensional example. 

We expect that a more general model with $p_i$ declining fast enough with $i$
is just a smeared version of this basic one, and its properties are not likely 
to differ from those of the simpler model. We conjecture that for our purposes, 
``fast enough'' decrease should be exponential with a rate logarithmically 
increasing in the dimension $n$, e.g. for large $k$,
\[
      p_k \le \exp(-\alpha(\log n)k),
\]
for some $\alpha>1$. (This is expected to be so because in this case the expected number of
neighbors of the focal genotype is finite.)

We observe that the first step of our procedure is an 
edge version of the percolation model discussed in the second section, with a 
similar giant component transition \citep{BKL1}. 
Namely, let $p_1=p_e=\lambda_e/n$. Then, if $\la_e>1$, there 
is a.~a.~s.~one giant conformist cluster of size $c\cdot 2^n$, with all others 
of size at most $Cn$. In contrast, if $\la_e<1$ all conformist clusters
are of size at most $Cn$. Note that the number of conformist 
clusters is always on the order $2^n$. In fact, even the number 
of ``non-conformist'' (i.e., isolated) clusters is a.~a.~s.~asymptotic to 
$e^{-\lambda_e} 2^n$, as $P(x\ \text{is isolated})=(1-\lambda_e/n)^n$. 

\begin{figure*}[t]
   \begin{center}
    {\includegraphics
    [clip, viewport= 140 325 475 680, height=4cm]{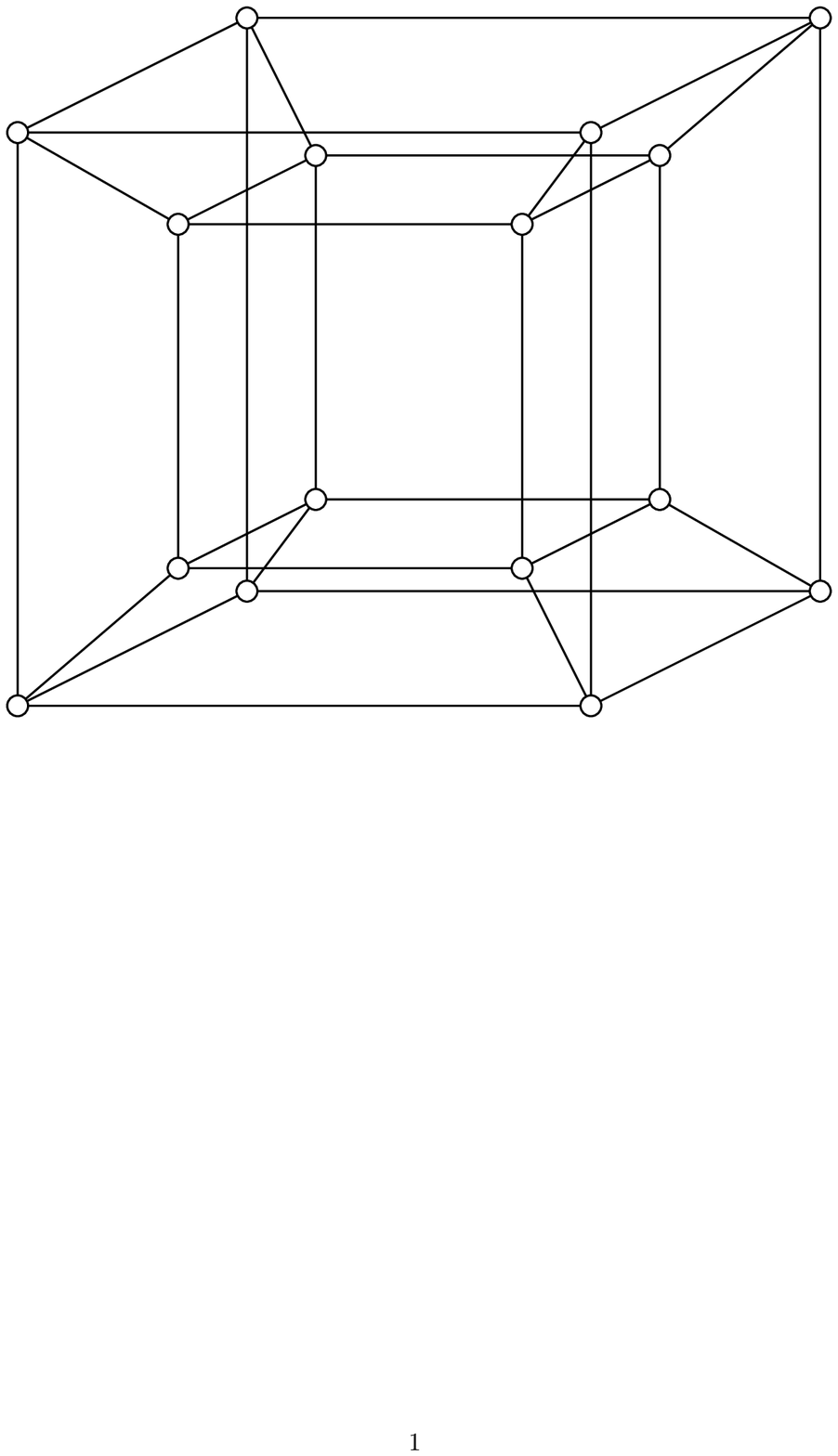}
    \hspace{1cm}
    \includegraphics[clip, viewport= 140 325 475 680, height=4cm]{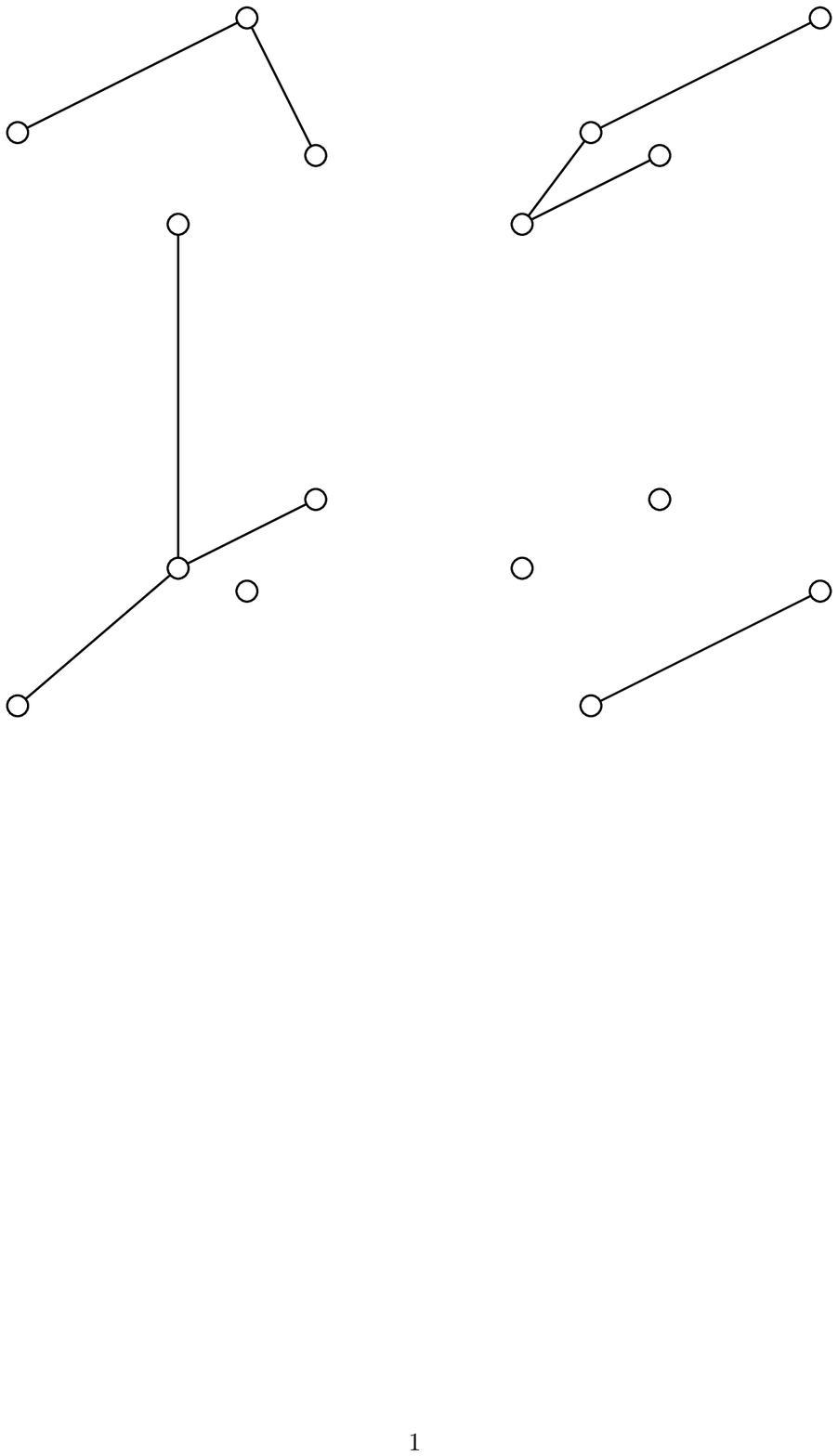} \\
    \vspace{.5cm}
    \includegraphics[clip,viewport= 140 325 475 680, height=4cm]{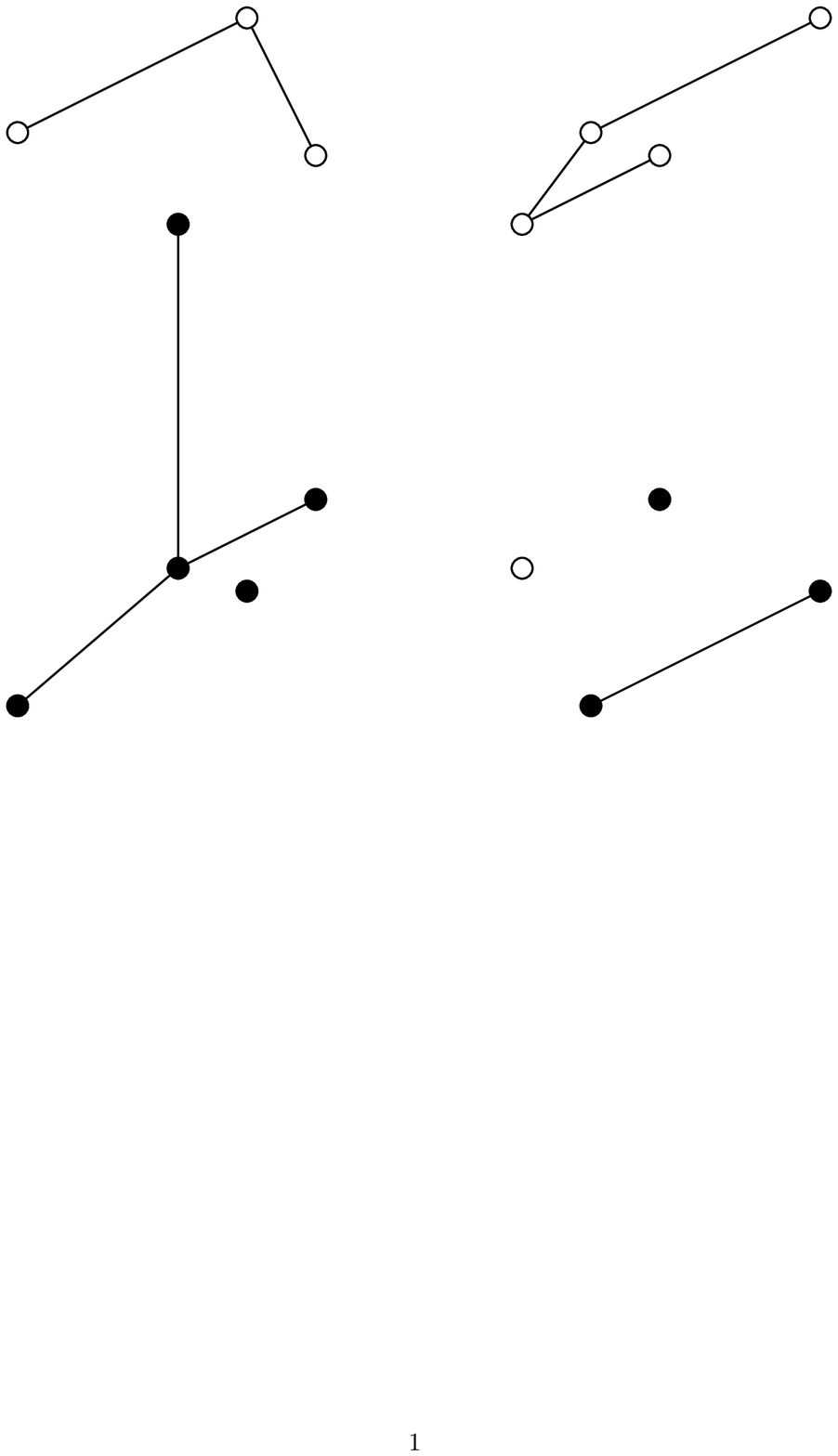}
    \hspace{1cm}
    \includegraphics[clip,viewport= 140 325 475 680, height=4cm]{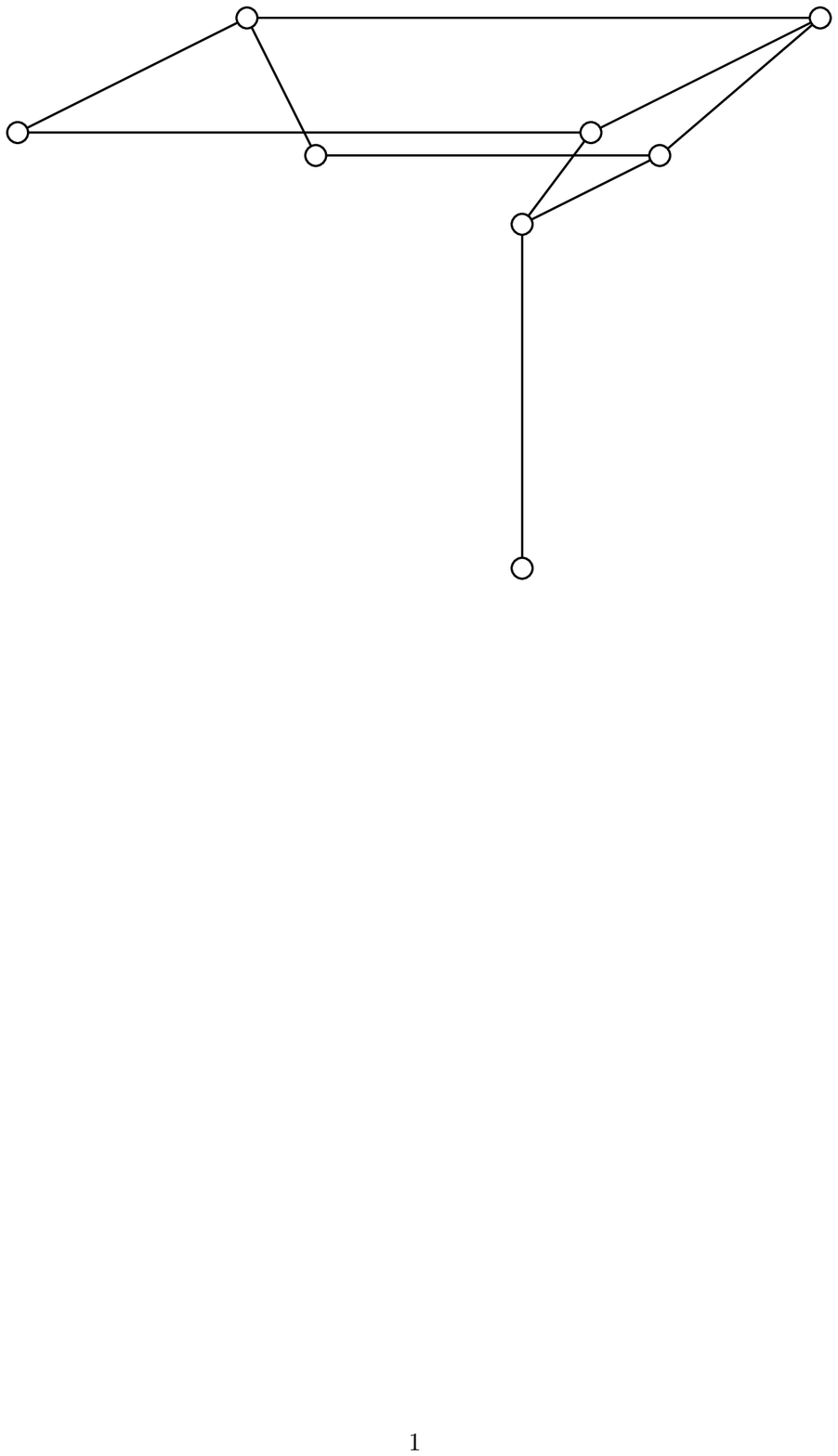}
    }\end{center}
\caption{A four-dimensional example: start with the cube $\cG^4$ (top left),  create conformist clusters by randomly eliminating each edge with probability 
 $1-p_e$ (top right), remove each conformist cluster with probability 
 $1-p_v$ (bottom left, removed vertices are black) and finally consider 
 connected components of the remaining vertices (bottom right, 
 there is just one component in this case).}
\end{figure*}

Denote by $x\conn y$ (resp.~$x\notconn y$) the event that 
$x$ and $y$ are (resp.~are not) in the same conformist cluster.
First, we note that the probability $P(x \conn y)$ that two genotypes belong to 
the same conformist cluster depends on the Hamming distance $d(x,y)$ between them, and on
$p_e=\lambda_e/n$. In particular,
we show in Appendix A that, if $\la_e<1$ and $d(x,y)=k$ is fixed, then 
\begin{equation} \label{Px-y}
k!p_e^k (1 - O(n^{-2})) \leq P(x \conn y) \leq k!p_e^k (1 + O(n^{-1} \log{n})). 
\end{equation}
The dominant contribution $k!p_e^k$ is simply the expected number of conformist pathways between $x$ and $y$ 
that are of shortest possible length. 

It is also important to note that, for every $x\in \cG$, 
the probability $P( x$ is viable$)=p_v$, therefore it does not depend on $p_e$. 
Moreover, for $x,y\in \cG$,  
$$
\begin{aligned} 
&P(x\text{ and }y\text{ viable})-p_v^2\\
&=P(x\text{ and }y\text{ viable},x\conn y)+ P(x\text{ and }y\text{ viable},x\notconn y)-p_v^2\\
&=p_vP(x\conn y)+ p_v^2\cdot P(x\notconn y)-p_v^2\\ 
&=p_v(1-p_v)P(x\conn y)\ge 0. 
\end{aligned} 
$$
Therefore, the correlation function~(\ref{rho}) is
\begin{equation}
\rho(x,y)=P(x\conn y), 
\end{equation} 
which clearly increases with $p_e$ and, thus, with $\lambda_e$. 
Therefore, this model 
has tunable positive correlations controlled by the parameter $\la_e$, whose value does 
not affect the expected number of viable genotypes. 
The correlation function $\rho(x,y)$ decreases exponentially with distance
$d(x,y)$ when $\la_e<1$, and is bounded below when $\la_e>1$.  Nevertheless, 
as we will see below, we can effectively use local methods for all values of $\la_e$. 

\subsection{Threshold surface for percolation} 

Proceeding by the local branching process heuristics, 
we reason that a surviving node on the branching tree can have 
two types of descendants: those that are connected by conformist mutations 
and those that are in different conformist clusters and thus
independently viable. Therefore the number 
of descendants is approximately Poisson($\la_e+\la_v$).
This can only work when $\la_e<1$, as otherwise the correlations are global.  

If $\la_e>1$, we need to eliminate the 
entire conformist giant component, which is \aas~inviable. 
Locally, we condition on the 
(supercritical) branching process of the supposed descendant to die out. 
Such conditioned process is a subcritical branching process, with 
Poisson $(\la_e\delta)$ distribution of successors \citep{AN}
where $\delta=\delta(\lambda_e)$ is given by the equation~(\ref{delta}). 
This gives the 
conformist contribution, to which we add the independent Poisson$(\la_v\delta)$ contribution. 

 \begin{figure*}[t]
  \begin{center}
  \vspace{5pt}
   \includegraphics[clip=true,height=10cm]{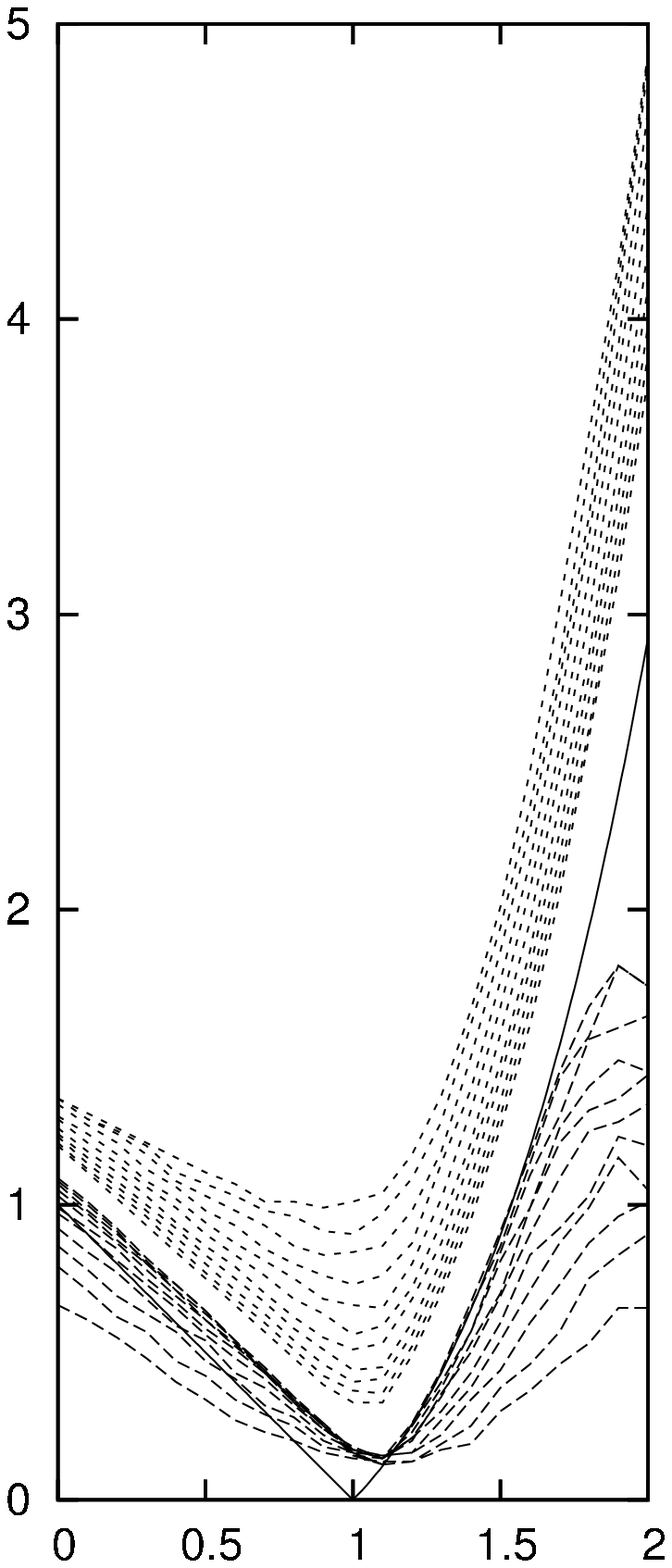} \hspace{1.5cm}
   \includegraphics[clip=true,height=10cm]{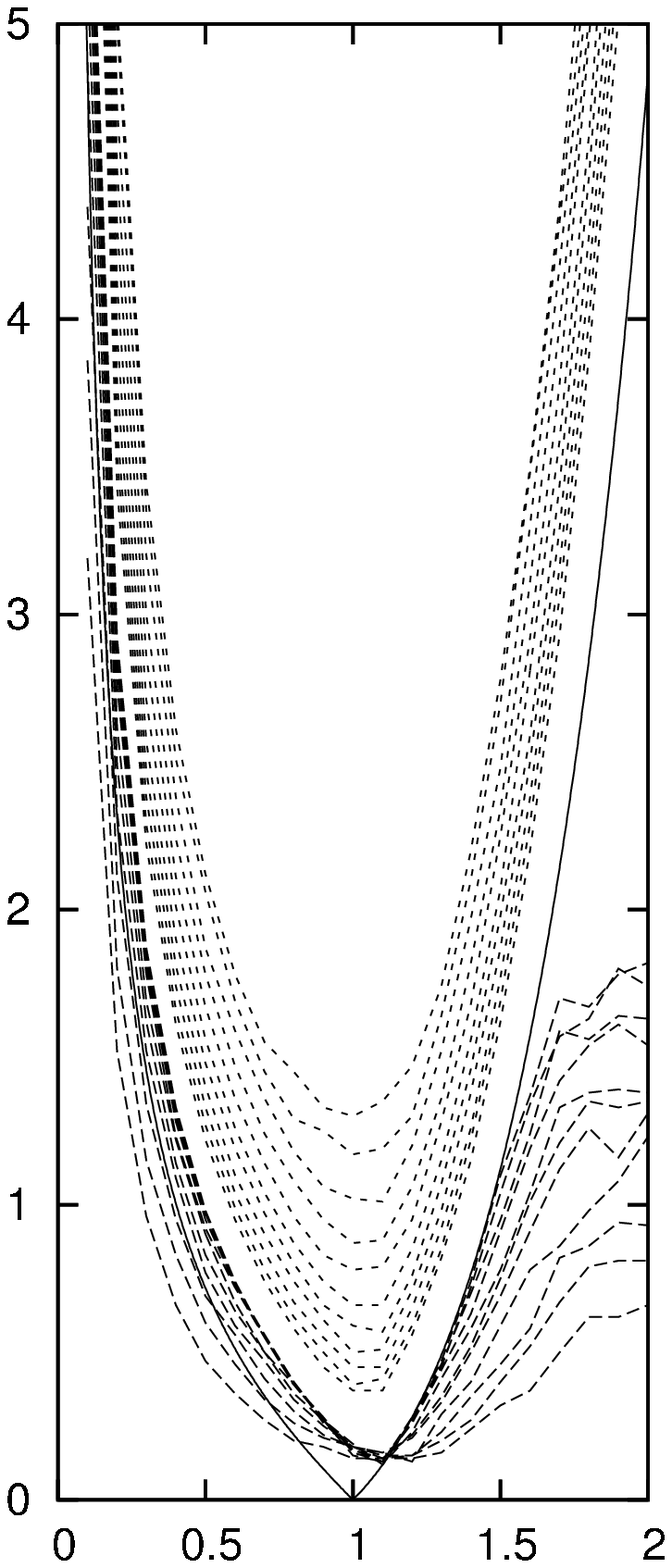}
  \end{center}
  
  \caption{Simulated $\la_v^m$ (long dashes) and  $\la_v^{M}$ (short dashes), and $\zeta$
  (solid) plotted against $\la_e$, for $n=10, \dots, 20$, and models from Section 
  4.1 (left frame) and Section 4.2 (right frame). Lower bounds increase with $n$, and 
  upper bounds decrease, for this range of $n$. } 
\end{figure*}

To have a convenient summary of the conclusions above, 
assume that $\la_e$ is fixed and let $\zeta(\la_e)$
be the smallest $\la_v$ which \aas~ensures the giant component, i.e., 
\[
\zeta(\la_e)=\inf\{\la_v: \text{a cluster of at least }cn^{-1} 2^n 
\text{ viable genotypes exists \aas~for some } c>0\}. 
\]  
One would expect that for $\la_v<\zeta(\la_e)$ all components are \aas~of size at most $Cn$. 
The asymptotic critical curve is given by 
$\la_v=\zeta(\la_e)$, where 
\begin{equation}  \label{pheno}
\zeta(\la)=
\begin{cases}
1-\la &\qquad\text{if } \la\in [0,1],\\
\frac 1{\delta}-\la&\qquad\text{if } \la\in [1,\infty). 
\end{cases}
\end{equation}  

Having only a heuristic proof of this, we resort to computer 
simulations for confirmation. For this, we   
indicate
global connectivity with the event $A$ that a genotype 
within distance 2 of $0\dots 0$ is connected 
(through viable genotypes) to a genotype 
within distance 2 of $1\dots 1$. 
We make this choice because the 
distance 2 is the smallest that works with asymptotic certainty. 
Indeed, the genotypes $0\dots0$ and $1\dots1$ are likely to be inviable. 
Even the number of viable genotypes within distance one of each of these is only of constant order, 
so even in the percolation regime the probability of connectivity between 
a viable genotype within distance one 
of $0\dots0$ and a viable one within distance one of $1\dots1$  does not converge to 1 but is of 
a nontrivial constant order. By contrast, there are about $n^2$ vertices 
within distance 2 of $0\dots0$ among which of order $n$ are viable. 
 
When $\la_v>\zeta(\la_e)$ the probability of the event $A$ 
should therefore be (exponentially) close to 1. On the other hand, when $\la_v<\zeta(\la_e)$ 
the probability that a connected component within distance 2 of either  
$0\dots0$ or $1\dots1$ extends for distance of the order $n$ 
is exponentially small. We further define the critical curves 
$$ 
\begin{aligned} 
&\text{$\la_v^{m}=\;$the smallest $\la_v$ for which 
$P(A)>0.1$,}\\ 
&\text{$\la_v^{M}=\;$the largest $\la_v$ for which 
$P(A)<0.9$.}  
\end{aligned}
$$

We approximated $\la_v^m$ and $\la_v^{M}$ for 
$n=10, \dots, 20$ and $\la_e=0(0.1)2$, with 1000 
independent realizations
of each choice of $n$, $\la_e$, and $\la_v$. We used the linear 
cluster algorithm described in \cite{Sed}. 
The results are depicted in Figure 1.
Unfortunately, simulations above $n\approx 20$ 
are not feasible.  

From Figure 2 we observe that:
\begin{itemize} 
 
\item Even for low $n$, both critical curves approximate well the 
overall shape of the theoretical limit curve $\zeta$. 
\item $\la_v^{m}$ and $\la_v^{M}$ get 
closer faster than they converge to $\zeta$. Consequently, 
one can expect that $P(A)$ makes a very sharp jump from near 0 
to near 1 even for moderate $n$. 
\item For $\la_e<1$, $\la_v^{m}$ tends to be above the limit curve. This is 
not really surprising, as the local argument always gives an upper 
bound on the probability $P(A)$ of event $A$. Further, the approximation of $\la_v^m$ deteriorates 
near $\la_e=2$, which stems from the possibility of survival of the 
giant component in this regime. 
\end{itemize}
 
What is clear from the heuristics and simulations is that 
conformist mutations, and thus correlations, significantly affect 
the probability of long range 
connectivity in the genotype space. The effect is not monotone:
the most advantageous choice 
is when the correlations are at the point of phase transition between between local and global.

To intuitively understand why percolation occurs the easiest with $\la_e \approx 1$, it helps
to think of the model as a branching process on clusters rather than on genotypes. 
For a genotype on a viable cluster, 
there is a number of neighboring clusters and each of these is viable with 
probability $p_v$. If $\lambda_e < 1$, then the probability that any two of the neighboring 
genotypes are in the same cluster is $o(1)$, so there are asymptotically exactly $n$ clusters 
neighboring the present cluster. Consequently, the overall number of descendants will be greater 
if the size of these clusters is greater on average; which is exactly what happens as $\lambda_e$ 
increases towards 1. If $\lambda_e > 1$, then there is a positive proportion of the neighboring 
genotypes that are in the giant cluster. This giant cluster is likely to be inviable, so the parameter 
$\lambda_v$ must be greater to compensate for its loss. 

\subsection{Correlations between conformity and viability} 

In the previous model, the viability probability $p_v$ 
was independent of the conformity structure. Mainly to 
investigate the robustness of our conclusions, 
we consider a simple generalization in which there 
are either positive or negative correlations between conformity 
and fitness. While more sophisticated models are possible, 
the one below is chosen for its amenability to relatively simple analysis.   

Assume now that conformist clusters are formed as before (i.e.,  
with edges being conformist with probability $p_e=\lambda_e/n$), 
are still independently viable, but 
now the probability of their viability depends on their 
size. We will consider the simple case when an isolated genotype 
(one might call it {\it non-conformist\/}) is viable with probability $p_0=\la_0/n$, 
while a conformist cluster of size larger than 1 is viable with probability $p_1=\la_1/n$.

In this case 
$$
P(x\text{ is viable})=(1-p_e)^np_0+(1-(1-p_e)^n)p_1\sim \frac 1n\left(   
e^{-\la_e}\la_0+(1-e^{-\la_e})\la_1\right).
$$
Moreover, by a similar calculation as before, 
$$
\begin{aligned}
&P(x\text{ and }y\text{ viable})-P(x\text{ viable})^2\\
&=p_1(1-p_1)P(x\conn y)+P(x\text{ non-conformist})^2p_e(p_0-p_1)^2\cdot 1_{\{d(x,y)=1\}}.
\end{aligned}
$$
Here, the last factor is  the indicator of the set $\{(x,y), d(x,y)=1\}$, which equals 
$1$ if $d(x,y)=1$ and $0$ otherwise.
Therefore, for $d(x,y)\ge 2$, the correlation function (\ref{rho})
is
$$
\rho(x,y)\sim\frac {\la_1}{e^{-\la_e}\la_0+(1-e^{-\la_e})\la_1}P(x\conn y), 
$$
which is smaller than before iff $\la_1<\la_0$. However, it has the same 
asymptotic properties unless $\la_1=0$.

Assume first that $\la_e<1$. 
The local analysis now leads to 
a {\it multi-type\/} branching process \citep{AN} with three types: NC (non-conformist node), 
CI (non-isolated node independently viable, so no conformist edge is 
accounted for), and CC (non-isolated node viable by conformity, so 
a conformist edge is accounted for).

Note first that a genotype is 
non-conformist with probability about $e^{-\la_e}$. 
Hence a node of any of the three types creates a Poisson($e^{-\la_e}\la_1$) number 
of type NC descendants, and a Poisson($(1-e^{-\la_e})\la_1$) number of type CI 
descendants. In addition, the type CI creates a Poisson($\la_e$), conditioned
on being nonzero, number of descendants of type CC and type CC creates a 
Poisson($\la_e$) number of descendants of type CC. Thus 
the matrix of expectations, in which the $ij$th entry is the expectation of the number 
of type $j$ descendants from type $i$, is 
\[
M=
\begin{bmatrix}
e^{-\la_e}\la_0 & \left(1- e^{-\la_e}\right)\la_1 & 0\\ 
e^{-\la_e}\la_0 & \left(1- e^{-\la_e}\right)\la_1 & \la_e/(1-e^{-\la_e})\\ 
e^{-\la_e}\la_0 & \left(1- e^{-\la_e}\right)\la_1 & \la_e \end{bmatrix}\quad . 
\]
When $\la_e>1$, $\la_e$ needs to be replaced by $\la_e\delta$, and 
$\la_1$ by $\la_1\delta$, where $\delta=\delta(\la_e)$ is given by ~(\ref{delta}).

It follows from the theory of multi-type branching processes \citep{AN} that 
the critical surface for survival of a multi-type 
branching process is given by $\det(M-1)=0$.
  
The simplest case is when only non-conformist genotypes may be viable, 
i.e., $\la_1=0$. In this case the critical surface is given by $\la_0 e^{-\la_e}=1$ (Pitman, unpub.).
Not surprisingly, the critical $\la_0$ to achieve global connectivity strictly 
increases with $\la_e$, which is the result of negative correlations between 
conformity and viability.

The other extreme is when non-conformist genotypes are inviable, 
i.e., $\la_0=0$. As an easy computation demonstrates, 
the critical curve is now given by $\la_1=\zeta(\la_e)$, where 
\begin{equation}\label{phenocorr}
\zeta(\la)=
\begin{cases}
\frac{1-\la}{\la e^{-\la}+1-e^{-\la}} &\qquad\text{if } \la\in \{0,1\},\\ 
\frac{\rho^{-1} -\la}{ \la e^{-\la}+1-e^{-\la\rho}}&\qquad\text{if } \la\in [1,\infty). 
\end{cases}
\end{equation}
Note that $\zeta(\la)\to \infty$ as $\la\to 0$. We carried out exactly the 
same simulations as before. These are also featured in Figure 2 (right frame), and again 
confirm our local heuristics. We conclude that positive correlations
between viability and conformity tend to lead to a V-shaped critical 
curve, whose sharpness at critical conformity $\la_e=1$ increases with 
the size of correlations. In short, then, correlations help more 
if viability probability increases with size of conformist clusters. 

\section{Percolation in incompatibility models}

In the model considered in the previous section 
correlations rapidly decreased with distance. This property
made local analysis possible. The models we introduce now 
are fundamentally different in the sense that correlations are 
so high that the local method gives a wrong answer. 

In the previous sections, in constructing fitness landscapes we were assigning fitness
to individual genotypes or phenotypes. Here, we make certain assumptions about ``fitness'' of
particular combinations of alleles or the values of phenotypic characters. Specifically,
we will assume that some of these combinations are ``incompatible'' in the sense that the
resulting genotypes or phenotypes have reduced (or zero) fitness \citep{orr95,orr96,gav04}.
The resulting models can be viewed as a generalization of the Bateson-Dobzhansky-Muller
model \citep{orr95,orr96,orr97,orr01,gav96b,gav97,gav97b,gav03d,gav04,coy04} 
which represents a canonical model of speciation.

\subsection{Diallelic loci} 

We begin by assuming that viability of a genotype is determined by 
a set $F$ of pairwise incompatibilities. $F$ is thus
a subset of  $4\cdot \binom{n}{2}$ pairs $(u_i, v_j)$, 
where $1\le i<j\le n$ and $u,v\in\{0,1\}$. In this nonstandard notation, $(0_1,0_2)\in F$, 
for example, means that allele $0$ at locus $1$ and allele $0$ at locus $2$ 
are incompatible. In general, if $(u_i, v_j)\in F$, 
all genotypes with $u$ in position $i$ and $v$ in position $j$ 
are inviable.  
A genotype $x$ is then inviable if and only if there exist $i$ and $j$, with $i<j$, 
so that $u$ and $v$ are, respectively, the alleles of $x$ at loci $i$ and $j$, 
and $(u_i, v_j)\in F$. 
For example, if $F_1=\{(0_1, 0_2), (1_2, 0_3), (1_1, 1_2)\}$, viable genotypes may have 
$011$, $100$, and $101$ as their first three alleles. For $F_2=F_1\cup \{(0_1, 1_3), (1_1, 0_2)\}$, 
no viable genotype remains. 

Incompatibility $(0_1, 0_2)$ is equivalent to two implications: $0_1\implies 1_2$ and 
$0_2\implies 1_1$ or to the single {\tt OR} statement $1_1$ {\tt OR} $1_2$. In this interpretation, 
the problem of whether, for a given list of incompatibilities $F$, there is a viable genotype is 
known as the {\tt $2$-SAT} problem \citep{KV}. 
The associated {\it digraph\/} $D_F$ is a graph on $2n$ vertices $x_i$, $i=1,\dots n$, $x=0,1$, 
with oriented edges determined by the implications. A well-known theorem \citep{KV} states 
that a viable genotype exists iff $D_F$ contains no oriented cycle 
from $0_i$ to $1_i$ and back to  $0_i$  for any $i=1,\dots n$ in $D_F$. 
For example, for the incompatibilities $F_2$ as above, 
one such cycle is $0_1\to1_2\to 1_3\to 1_1\to 1_2\to 0_1$. 

Now assume that each possible incompatibility is adjoined to $F$ at random, independently 
with probability 
\[
p=\frac c{2n}.
\]
(We use the generic notation $p$ for a probability parameter
in all our models, even though the nature of probabilistic assignments differs from model to model.)

{\bf Existence of viable genotypes.}\quad
Let $N$ be the number of viable genotypes. Then
\begin{itemize} 
\item if $c>1$, then a.~a.~s.~$N=0$. 
\item if $c<1$, then a.~a.~s.~$N>0$.
\end{itemize}
This result first appeared in the computer science literature in the 90's
(see \citealt{dlV} for a review), and it is an 
extension of the celebrated Erd\"os-R\'enyi random graph results 
\citep{Bol,JLR} to the oriented case. 

Note that the expectation
$E(N)=2^n(1-p)^{\binom{ n}{2}}\approx 2^ne^{-cn/4}$, 
which grows exponentially whenever $c<4\log 2\approx 2.77$. Neglecting 
correlations would therefore suggest a wrong threshold for $N>0$. The local method 
(e.g., used in \citealt[Chapter 6]{gav04}) is 
even farther off, as it suggests an \aas~giant component when $p<(1-\e)\log n/n$
for any $\e>0$. 

{\bf The number of viable genotypes.}\quad
Assume that $c<1$. Sophisticated, but not mathematically rigorous 
methods based on {\it replica symmetry\/} \citep{MZ,BMW}  from statistical physics suggest that, 
as $n\to\infty$, 
$\lim n^{-1}\log N$ varies almost linearly between 
$\log 2\approx 0.69$ (for small $c$, when, as we prove below, this limit is 
$\log 2+\cO(c)$) and about $0.38$ (for $c$ close to $1$). 
One can however prove that $n^{-1}\log N$ is for large $n$ sharply 
concentrated around its mean \citep{dlV}.

Upper and lower bounds on $N$ can also be obtained 
rigorously. For example, if $X$ is a number of 
incompatibilities which involve {\it disjoint\/} pairs of loci
(i.e., those for which every locus is represented at most once among the 
incompatibilities), 
then $N\le \exp(n\log 2+X\log(3/4))$, as each of the $X$ incompatibilities 
reduces the number of viable genotypes by the factor $3/4$.  
If we imagine 
adding incompatibilities one by one at random until 
there are about $cn$ of them, then after we have $k$ 
incompatibilities on disjoint pairs of loci the waiting time (measured by
the number of incompatibilities added) 
for a new disjoint one is geometric with expectation $\binom{n} {2}/\binom{n-2k} {2}$. 
Therefore, 
$X$ is \aas~at least $Kn$, where 
$K$ solves the approximate equation 
$$
\binom{n} {2} \left(\sum_{k=0}^{Kn}\frac 1{ \binom{n-2k} {2}} \right)\sim cn, 
$$
or 
$$
\int_{0}^{Kn}\frac 1{(n-2k)^2}\, dk \sim \frac cn, 
$$
which reduces to $K=c/(1+2c)$. This implies that the upper bound on $N$ can be 
defined as
\begin{equation}  \label{up_bound}
\limsup \frac 1n\log N\le \frac {1}{1+2c}\log 2+\frac {c}{1+2c}\log 3.
\end{equation} 

A lower bound is even easier to obtain. Namely, 
the probability that a fixed location (i.e., locus) $i$ does not appear in $F$ is $(1-p)^{4(n-1)}
\to e^{-2c}$, and then it is easy to see that the number of loci represented in $F$ 
is asymptotically $(1-e^{-2c})n$. As the other loci are neutral (in the sense that changing
their alleles does not affect fitness), 
$n^{-1}\log N$ is asymptotically at least $e^{-2c}\log 2$. Clearly, this gives
a lower bound on the exponential size of any cluster of viable genotypes. 

If this was an accurate bound, it would imply that the space of 
genotypes is rather simple, in that almost all its entropy would come from neutral loci. The Appendix B presents two arguments which will 
demonstrate that this is not the case. The derivations there are somewhat technical, 
but do provide more insight into random pair incompatibilities. 

{\bf The structure of clusters.}\quad
The derivations in Appendix B show that every viable genotype is connected 
through mutation to a fairly substantial 
viable sub-cube. In this sub-cube, alleles on at most a proportion $r_u(c)<1$ of loci 
are fixed (to 0 or 1) while the remaining proportion $1-r_u(c)$ could be 
varied without effect on fitness. Note from Figure 4 in the 
Appendix B that $1-r_u(c)\ge 0.3$ for 
all $c$, and that such a phenomenon is 
extremely unlikely on uncorrelated landscapes.
Note also that, for $c<1$, $N\ge 2^{(1-r_u(c))n}$ \aas~and so  the lower
bound on $N$ can be written as
\begin{equation} \label{low_bound}
\liminf\frac 1n\log N\ge (1-r_u(c))\log 2. 
\end{equation} 

{\bf The number of clusters.}\quad
The natural next question concerns the number of clusters
$R$ when $c<1$. This again has quite a surprising answer, unparalleled in 
landscapes with rapidly decaying correlations. Namely, 
$R$ is {\it stochastically bounded\/}, that 
is, for every $\e>0$ there exists an $z=z(\e)$ such that $P(R\le z$ for all $n)>1-\e$.
As there is some confusion in the literature as to whether it is even possible 
to get more than one cluster \citep{BMW}, Appendix C
presents a sketch of the results which will appear in Pitman (unpub.).
There we also show that the limiting probability of a unique cluster is
$\sqrt{(1-c)e^c}$.

Asymptotically, a unique cluster has a better than even chance of 
occurring for $c$ below about $0.9$, and is {\it very\/} likely to occur 
for small $c$, though of course not 
\aas~so. To confirm, we have done simulations for $n=20$ and $c=0.01 (0.01) 1$
(again 1000 trials in each case) and got distribution of clusters depicted 
in Figure~3. The results suggest that the convergence to limiting distribution 
is rather slow for $c$ close to 1, and that the likelihood of a unique 
cluster increases for low $n$. 

\begin{figure*}[t]
  \begin{center}
   {\includegraphics[clip=true,height=5cm]{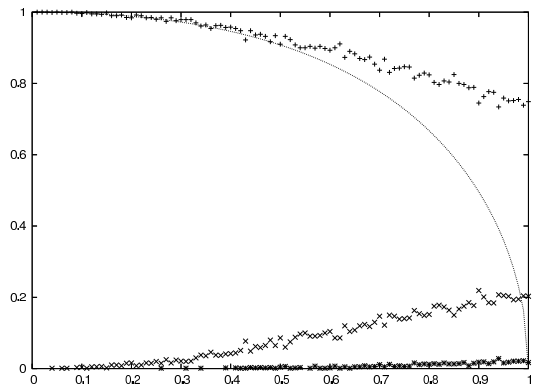}
    }
  \end{center}
  
  \caption{Simulated number of clusters, vs. $c$ for $n=20$. The
  proportion (out of 1000) of trials with exactly one, exactly two, and at least three clusters 
  is plotted respectively with $+$'s, $\times$'s and $*$'s. The solid curve is 
  $\sqrt{(1-c)e^c}$. 
  } 
\label{number_clusters}
\end{figure*}

To summarize, in the presence of random pairwise incompatibilities, the set 
of viable genotypes is, when nonempty, 
divided into a stochastically bounded number of connected clusters, 
where a unique cluster is usually the most likely possibility. 
These clusters are all of exponentially large size
(with bounds given by equations \ref{up_bound} and \ref{low_bound}), in fact they all contain 
sub-cubes of dimension at least $(1-r_u(c))n$. 
However, the proportion 
of viable genotypes among all $2^n$ genotypes is exponentially small, by 
equation (\ref{up_bound}). 

\subsection{Multiallelic loci} 

Here we assume that at each locus there can be $a\ (\ge 2)$
alleles (cf., \citealt{Rei}). In this case, the genotype space is
 the generalized hypercube
$\cG_a=\{0,\dots, a-1\}^n$. For $a=3$ 
this could be interpreted as the genotype space of diploid 
organisms without {\it cis-trans\/} effects \citep{gav97b},
$a=4$ corresponds to DNA sequences, and $a=20$ corresponds to proteins. 
Much larger values of $a$ can correspond to a number of alleles at a protein 
coding locus and we will see later that 
there is not much difference between this model and a
natural continuous space model. 

We will assume that each pair of alleles, out of total 
number of $a^2\binom{n}{2}$ is independently incompatible
with probability 
$$p=\frac{c}{2n}.$$ 
The main question we are interested in 
here is for which values of $c$ viable genotypes exist {\aas }

Clearly, if $N$ is the number of viable phenotypes, then the expectation
$$
E(N)=a^n(1-p)^{\binom{n}{2}}\approx\exp(n \log a-{\textstyle\frac 14}cn), 
$$
and so there are \aas~no viable phenotypes when $c>4\log a$. On the 
other hand, clearly there are viable genotypes 
(with all positions filled by 0's and 1's) when $c<1$. It turns out that the 
first 
of these trivial bounds is much closer to the critical value when $a$ 
is large. Before we proceed, however, we state a sharp 
threshold result from \cite{Mol}: there exists a function $\gamma=\gamma(n,a)$ 
so that for every $\e>0$, 
\begin{itemize} 
\item if $c>\gamma+\e$, then a.~a.~s.~$N=0$. 
\item if $c<\gamma-\e$, then a.~a.~s.~$N>0$.
\end{itemize}
In words, for a fixed $a$, the probability of the event that $N\ge 1$ 
transitions sharply from large to small
as $np$ varies. As it is not proved that 
$\lim_{n\to\infty}\gamma(n,a)$ exists, it is in principle possible 
that the place of this sharp transition fluctuates as $n$ increases 
(although it must of course remain within $[1, 4\log a]$). 

Our main result in this section is 
\begin{equation}\label{gamma}
\gamma=4\log a-o(1), \text{ as }a\to\infty. 
\end{equation}
This somewhat surprising result in proven in Appendix D by the 
second moment method, as developed in \cite{AM} and \cite{AP}. 

\subsection{Continuous phenotype spaces} 

Here we extend the model of pair incompatibilities for the case of continuous 
phenotypic space $\cal{P}$. Again, we have a small $r>0$ as a parameter. 
For each of $(i,j)$, $i<j$, we consider independent Poisson point location $\Pi_{ij}$ 
in the unit square $[0,1]\times[0,1]$, of rate $\la=c/(2n)$. (Equivalently, choose Poisson($\la$) number of 
points uniformly at random in $[0,1]\times[0,1]$.) Then we declare $a\in \cP$ inviable 
if there exist $i<j$ so that $(a_i,a_j)$ is within $r$ of $\Pi_{ij}$. 
Again, we use the two-dimensional $\ell^\infty$ norm for distance. 
Our procedure can be visualized as throwing a random number of
$(n-2)$-dimensional square tubes of inviable phenotypes into the phenotype space.

Our main result here is that the existence threshold is on the order $c\approx -\log r/r^{2}$. 
Namely, we prove in the Appendix E that there exists a constant $C>0$ so that for small enough $r$, 

 \begin{itemize} 
\item if $c>4\frac{-\log r}{r^2}$, then a.~a.~s.~$N=0$. 
\item if $c<\frac{-\log r-C}{r^2}$, then a.~a.~s.~$N>0$.
\end{itemize}

\subsection{Complex incompatibilities} 

Here we assume that incompatibilities involve $K\ (\geq 2)$ diallelic loci \citep{orr96,gav04}.
The question whether a viable combination of genes exist is then equivalent to 
the {\tt $K$-SAT} problem \citep{KV}. Even for $K=3$, this is an NP-complete problem 
\citep{KV}, so there is no known polynomial algorithm to answer this question.
The random case, which we now describe, is also much harder to analyze 
than the {\tt $2$-SAT} one. 
Let $F$ be a random set 
to which any of the $2^K\binom n K$ incompatibilities belong independently with 
probability 
$$
p=\frac {K!}{2^K}\cdot \frac c{n^{K-1}}.
$$
Here $c=c(K)$ is a constant, and the above form has been 
chosen to make the number of incompatibilities in $F$ asymptotically $cn$.
(Note also the agreement with the definition of $p$ in Section 5.1 
when $K=2$.)  For a fixed $K$,  
it has been proved \citep{Fri} that the probability that viable genotype exists
jumps sharply from 0 to 1 as $c$ varies. However, the location of the 
jump has not been proved to converge as $n\to\infty$. Instead, 
a lot of effort has been 
invested in obtaining good bounds. For example \citep{AP}, for $K=3$, $c<3.42$ implies {\aas } 
existence 
of viable genotype, while $c>4.51$ implies \aas~nonexistence (while the sharp 
constant 
is estimated to be about $4.48$, see e.g. \citealt{BMW}). 
For $K=4$ the best current bounds are $7.91$ and $10.23$. For large 
$K$, the transition occurs at $c=2^K\log 2-\cO(K)$ \citep{AP}. 

Techniques from statistical physics \citep{BMW} strongly suggest 
that, for $K\ge 3$, there is another phase transition, which 
for $K=3$ occurs at about $c=3.96$. For smaller $c$, the  
viable genotypes are conjectured to 
be contained in a {\it single\/} cluster. 
For larger $c$, the space of viable genotypes
(if nonempty) is divided into exponentially many connected clusters.

Perhaps more relevant to genetic incompatibilities is the following 
{\it mixed\/} model (commonly known as {\tt $(2+p)$-SAT}), \citealt{MZ}). Assume that 
every 2-incompatibility is present with probability $c_2/(2n)$, 
while every 3-incompatibility is present with probability $3c_3/(4n^2)$. 
The normalizations are chosen so that the numbers of the two types of
incompatibilities are asymptotically $c_2 n$ and $c_3 n$, respectively. 

If $c_2$ (resp. $c_3$) is very small, then the respective incompatibility 
set affects a very small proportion of loci, therefore 
$c_3$ (resp. $c_2$) determines whether a viable genotype is likely to exist. 
Intuitively, one also expects that 2-incompatibilities should be more 
important than 3-incompatibilities 
as one of the former type excludes more genotypes than one of the latter type.  A careful
analysis confirms this. First observe  
that $c_2>1$ implies \aas~non-existence of a viable genotype. The surprise
\citep{MZ,AKKK} is that if $c_3$ is small enough, $c_2<1$ 
implies \aas~existence of viable genotypes, so the 3-incompatibilities 
do not change the threshold. This is established in \cite{MZ} by a physics argument 
for $c_3<0.703$, while 
\cite{AKKK} gives a rigorous argument for $c_3<2/3$. Therefore, even if their numbers are 
on the same scale, if the more 
complex incompatibilities are rare enough compared to the pairwise 
ones, their contribution to the structure of the space of 
viable genotypes is not essential.

\section{Notes on neutral clusters in the discrete {\it NK\/} model} 

The model considered here is a special case of the discretized NK model \citep{kau93}, 
introduced in \cite{NE}. 
This model features $n$ diallelic loci each of which interacts with $K$ other loci.
To have a concrete example, assume that the loci are arranged on a
circle, so that $n+1\equiv 1$, $n+1\equiv 2$, etc., and let the 
interaction {\it neighborhood\/} of the $i$'th locus consist of itself 
and $K$ loci to its right $i+1, \dots, i+K$. For a given 
genotype $x\in\cG=\{0,1\}^n$,  
the neighborhood configuration of the 
$i$'th locus is then given by $\cN_i(x)= (x_i, x_{i+1}, \dots, x_{i+K})\in \{0,1\}^{K+1}$. 
To each locus and to each possible configuration
in its neighborhood 
we independently assign a binary fitness contibution. 
To be more precise, 
we choose the $2^{K+1}n$ numbers $v_i(y)$, $i=1, \dots, n$ and $y\in \{0,1\}^{K+1}$,  
to be independently 0 or 1 with equal probability, and interpret $v_i(y)$ 
as the fitness contribution of locus $i$ when its neighborhood configuration 
is $y$. The fitness
of a genotype $x$ is then the sum of contributions from each locus: 
$$
w(x)=\sum_{i=1}^n v_i(\cN_i(x)). 
$$
In \cite{kau93}, the values $v_i$ were taken from a continuous distribution.
In \cite{NE}, these values were integers in the range $[0,F-1]$ so that our model
is a special case $F=2$.
{\it Neutral clusters\/} are connected components of same 
fitness.

The $K=0$ case is easy but nevertheless illustrative.
Namely, a mutation at locus $i$ will not change fitness iff  
$v_i(0)=v_i(1)$; let $D$ be the number of such loci. 
Then $D\sim n/2$ \aas, the number of different fitnesses is $n-D$,  
each neutral cluster is a sub-cube 
of dimension $D$, and there are exactly $2^{n-D}$ neutral 
clusters. 

The next simplest situation is when $K=1$. Let 
$D_1$ be the number of loci $i$ 
for which $v_i$ is constant. Then 
$D_1\sim n/8$ \aas, and each neutral cluster contains a 
sub-cube of dimension $D_1$. Moreover, let $D_2$ be 
the number of loci $i$ for which $v_i(00)=v_i(01)\ne v_i(10)=v_1(11)$. 
Note that any genotypes that differ at such locus $i$ must belong to 
a different neutral cluster, and so the number 
of different neutral clusters is at least $2^{D_2}$. Thus there 
are exponentially many of them, as 
again $D_2\sim n/8$ {\aas }
This division of genotype space into exponentially many clusters
of exponential size persists for every $K$, although 
the distribution of numbers and sizes of these clusters is not well understood (see 
\citealt{NE} for simulations for $n=20$). 
 
Finally, we mention that the question of whether a 
genotype with the maximal possible fitness $n$ 
exists for a given $K$ is in many way related to issues in incompatibilities models
\citep{CJK}. 

\section{Discussion}

In this section we summarize our major findings and provide their biological interpretation.

The previous work on neutral and nearly neutral networks in multidimensional fitness
landscapes has concentrated exclusively on genotype spaces in which each individual
(or a group of individuals) is characterized by a discrete set of genes. However
many features of biological organisms that are actually observable and/or measurable are described by
continuously varying variables such as size, weight, color, or concentration. A question
of particular biological interest is whether (nearly) neutral networks are as prominent
in a continuous phenotype space as they are in the discrete genotype space. Our results
provide an affirmative answer to this question. Specifically, we have shown that in a simple
model of random fitness assignment, viable phenotypes are likely to form a large connected
cluster even if their overall frequency is very low provided the dimensionality of the phenotype
space, $n$, is sufficiently large. In fact, the percolation threshold for the probability
of being viable scales with $n$ as $1/2^n$ and, thus, decreases much faster than $1/n$ which is 
characteristic of the analogous discrete genotype space model.

Earlier work on nearly neutral networks has been limited to consideration of the relationship
between genotype and fitness. Any phenotypic properties that usually mediate this relationship
in real biological organisms have been neglected. In Section 4, we proposed a novel model in which 
phenotype is introduced explicitly. In our model, the relationships both between genotype and 
phenotype and between phenotype and fitness are of many-to-one type, so that neutrality is present
at both the phenotype and fitness levels. Moreover, this model results in a correlated fitness 
landscape in which the correlation function can be found explicitly. We studied the effects
of phenotypic neutrality and correlation between fitnesses on the percolation threshold and 
showed that the most conducive
conditions for the formation of the giant component is when the correlations are at the point
of phase transition between local and global. 
To explore the robustness of our conclusions, we then look at a simplistic but
mathematically illuminating model in which there is a correlation between conformity (i.e.,
phenotypic neutrality) and fitness. The model has supported our conclusions.

Section 5, we studied a number of models that have been recently proposed
and explored within the context of studying speciation. In these models, fitness is assigned to 
particular gene/trait combinations and the fitness of the whole organisms depends on the presence
or absence of incompatible combinations of genes or traits.  In these models, the correlations
of fitnesses are so high that local methods lead to wrong conclusions.
First, we established the connection between these models and $K$-{\tt SAT} problems, prominent
in computer science. Then we analyzed the conditions for the existence of viable genotypes,
their number, as well as the structure and the number of clusters of viable genotypes.
These questions have not been studied previously. Among other things we showed that the number
of clusters is stochastically bounded and each cluster contains a very large sub-cube.
The majority of our results are for the case of pairwise incompatibilities between diallelic
loci, but we also looked at multiple alleles and complex incompatibilities.  Moreover, we generalized
some of our results to continuous phenotype spaces.

At the end, we provided some additional results on the size, number and structure of
neutral clusters in the discrete $NK$ model.

Some more general lessons of our work are that
\begin{itemize} 
\item Correlations may help or hinder connectivity in fitness landscapes. Even  when 
correlations are positive and tunable by a single 
parameter, it may be advantageous 
(for higher connectivity) to increase 
them only to a limited extent. 
\item Averages (i.e., expected values) can easily lead to wrong conclusions, 
especially when correlations are strong. Nevertheless, they may still 
be useful with a crafty choice of relevant statistics.  
\item Very high correlations may fundamentally change the structure of connected 
clusters. For example, clusters may look locally more like cubes than trees and 
their number may be reduced dramatically.   
\item Necessary analytical techniques may be unexpected and quite sophisticated; 
for example, they may require
detailed understanding of random graphs, spin-glass machinery, or decision algorithms. 
\end{itemize}

{\small ACKNOWLEDGMENTS. 
This work was supported by the Defense Advanced Research Projects Agency (DARPA), 
by National Institutes of Health (grant GM56693),
by the National Science Foundation (grants DMS-0204376 and DMS-0135345), 
and by Republic of Slovenia's Ministry of Science (program P1-285).}

\newpage  

\section*{Appendix}

\subsection*{Appendix A. Proof of equation~(\ref{Px-y}).}

To prove equation (5), we assume that $\lambda_e<1$ and 
show that for a fixed $k$ (which does not grow with $n$), the
event that $x$ and $y$ at distance $k$ are in the same conformist cluster is most likely to
occur because $x$ and $y$ are connected via the shortest possible path. Indeed, 
the dominant term $k!p_e^k$ is the expected number of conformist pathways between $x$ and $y$ 
that are of shortest possible length $k$. This easily follows from the observation that 
on a shortest path there 
is no opportunity to backtrack; each mutation must be toward the other genotype. 
We can assume that $x$ 
is the all 0's genotype and $y$ is the genotype with 1's in the 
first $k$ positions and 0's elsewhere. 
There are $k!$ orders in which the 1's can be added.

To obtain the lower bound we use inclusion-exclusion on the probability 
that $x \conn y$ through a shortest path. Let $\mathcal{I}_l=\mathcal{I}_l(x,y)$ 
be the set of all paths of length $l$ between $x$ and $y$. 
Then 
$$P(x \conn y) \geq \sum_{\alpha \in \mathcal{I}_k} P(A_\alpha) -
\sum_{\alpha \neq \beta \in \mathcal{I}_k} P(A_\alpha\cap A_\beta)$$
where $A_\alpha$ is the event that a particular path $\alpha$ consists entirely 
of conformist edges.
Notice that two distinct paths of the same length differ by at least two edges. 
Thus, we get the following upper bound
$$\sum_{\alpha, \beta} P(A_\alpha\cap A_\beta) < (k!)2 p_e^{k+2},$$
and the lower bound in (5) follows.

The upper bound is a little more difficult to obtain (it is only here 
that we use $\lambda_e<1$) and we need some notation. 
Each genotype can be identified with the set of 1's that it contains, 
so for any two genotypes $u$ and $v$ we let $u \bigtriangleup v$ denote the set 
of loci on which they differ. Notice that if $u \bigtriangleup v$ 
is even (resp. odd) then every path between $u$ and $v$ is of even (resp. odd) 
length because each mutation which alters the allele at a locus not in $u \bigtriangleup v$ 
must later be compensated for.

To estimate the expected number of conformist pathways, 
we will need to bound the number of paths of length $l$ between $x$ and $y$. This is given by
$$ k!\binom{l}{m}m!n^{m}\quad \text{ where }\quad m=\frac{l-k}{2}.$$
We show this via the methods of \cite{BKL1}. 
They obtain an estimate for the number of cycles of a given length through a fixed vertex of the cube.

Given a path, say $x=v_0,v_1,\ldots,v_l=y$, between $x$ and $y$,
let us associate the sequence
$(\epsilon_1i_1,\ldots,\epsilon_l i_l)$
where
$$v_j \bigtriangleup v_{j-1}=\{i_j\}
\quad\text{and}\quad
\epsilon_j=
\left\{
\begin{array}{l}
+1\qquad\text{ if } v_j=v_{j-1}\cup{i_j} \\
-1\qquad\text{ if } v_j=v_{j-1}\setminus\{i_j\}
\end{array}
\right.$$
$j=1,\ldots,l$. Since distinct paths will have distinct sequences we
can bound the number of paths by finding an upper bound for the
number of sequences.

Note that there must be $m+k$ positive entries, which occur at
$\binom{l}{m+k}=\binom{l}{m}$ possible locations. The absolute
values of $m$ of these entries are chosen freely from $\{1,\dots, n\}$, while
the remaining $k$ must be the integers $1,\ldots,k$. There are
$n^mk!$ ways to do this. We are free to order the $m$ negative
entries and the bound follows.

We now assume that $d(x,y)$ is even and relabel $d(x,y)=2k$. 
We omit the similar calculation for odd distances. Define
$b=-3k/(2\log\la_e)$ and $t=\lfloor b\log  n\rfloor$. Then the 
expected number of conformist paths between $x$ and $y$ can be expressed as
\begin{eqnarray*}\sum_{l\geq k+1} \sum_{\mathcal{I}_{2l}} p_e^{2l}&=&
\sum_{k+1\leq l< t}
\sum_{\mathcal{I}_{2l}}p_e^{2l}+\sum_{l\geq t}
\sum_{\mathcal{I}_{2l}}p_e^{2l} \\
&<&\sum_{k+1\leq l< t} \binom{2l}{l-k}n^{l-k}(l-k)!(2k)!p_e^{2l}
+\sum_{l\geq t}n^{2l}p_e^{2l} \\
&=&\sum_{k+1\leq l< t}
(2l)^{l-k}n^{l-k}p_e^{2(l-k)}(2k)!p_e^{2k}
+\sum_{l\geq t}\la_e^{2l} \\
&<&(2k)!p_e^{2k}\sum_{l\geq k+1}(2b\la_e p_e\log n)^{l-k}+O(\la_e^{2b\log n})
\\
&=&k (2k)!p_e^{2k} O(p_e\log{n})+O(n^{2b\log \la_e}) \\
&=&k (2k)!p_e^{2k} O\left( n^{-1} \log{n} \right)  .
\end{eqnarray*}
 
\subsection*{Appendix B. Cluster structure under random pair incompatibilities.}

Here we show that, under random pairwise incompatibilities model introduced in Section 5.1,
connected clusters include large subcubes. The basic idea
comes from \cite{BD}. A configuration $a\in \{0,1,*\}^n$ 
is a way to specify a sub-cube of $\cG$, if $*$'s are thought of as places which could be filled 
by either a 0 or a 1. The number of non-$*$'s is the {\it length\/} of $a$. Call $a$ 
an {\it implicant\/} if the entire sub-cube specified by $a$ is viable.

We present two arguments, beginning with the one which 
works better for small $c$. Let the auxiliary random 
variable $X$ be the number of pairs of loci $(i,j)$, $i<j$, for which:  
\begin{itemize}
\item[(E1)] There is exactly one incompatibility involving alleles on $i$ and $j$.
\item[(E2)] There is no incompatibility involving an allele on either $i$ or $j$, 
and  an allele on $k\notin\{i,j\}$.  
\end{itemize}
Assume, without loss of generality, that the incompatibility 
which satisfies (E1) is $(1_i, 1_j)$. Then fitness of all
genotypes which have any of the allele assignments $0_i0_j$, $0_i1_j$ and $1_i0_j$, 
and agree on other loci, is the same. 
Note also that all pairs of loci which satisfy (E1) and (E2) must be 
disjoint. 
Therefore, if $x$ is any viable genotype, its cluster contains 
an implicant  with the number of $*$'s at least $X$ plus the number 
of free loci. To determine the size of $X$, note that the expectation
$$
E(X)={\binom{n}{2}}4p(1-p)^3(1-p)^{8(n-2)}\sim ce^{-4c}n
$$
and furthermore, by an equally easy computation, 
$$
E(X^2)-E(X)^2=\cO(n), 
$$
so that $X\sim ce^{-4c}n$ {\aas } 
It follows that every cluster 
contains \aas~at least $\exp((e^{-2c}+ce^{-4c})\log 2-\e)n)$, 
viable genotypes, for any $\e>0$.

The second argument is a refinement of the one in \cite{BD} 
and only works better for larger $c$. 
Call an implicant $a$ a 
{\it prime implicant (PI)\/} if at any locus 
$i$, replacement of either $0_i$ or $1_i$  
by $*_i$ results in a non-implicant. Moreover, we call $a$ the {\it least prime 
implicant (LPI)\/} if it is a PI, and the following two conditions are 
satisfied. First, if all the $*$'s 
are changed to 0's, then  no change from $1_i$ to $0_i$ results in a
viable genotype.
Second, 
no change $*_i1_j$ to $1_i*_j$, where $i<j$, results in an indicator. 

Now, every viable genotype must have an LPI in its cluster.
To see this, assume we have a PI for which the first condition is not satisfied. Make the 
indicated change, then replace some 0's and 1's by $*$'s 
until you get a prime indicator. If the second 
condition is violated, make the resulting switch, then again 
make some replacement by $*$'s until you arrive at a PI. 
Either of these two operations moves within the same cluster, and 
keeps the number of 1's nonincreasing 
and their positions more to the left. Therefore, the procedure 
must at some point end, resulting in an LPI in the same cluster.  

For a sub-cube $a$ to be an LPI,
the following conditions need to be satisfied: 
\begin{itemize}
\item[(I1)] Every non-$*$ has to be compatible with every other non-$*$, 
and with both 0 and 1 on each of the $*$'s. 
\item[(I2)] Any of the four 0,1 combinations on any pair of $*$'s must be compatible.
\item[(LPI1)] Pick an $i$ with allele 1, that is, a $1_i$. 
Then $0_i$ must be incompatible with at least 
one non-$*$, or at least one 0 on a $*$. Furthermore, if $0_i$ has an 
incompatibility
with a 0 on a $*$ to its left, it has to have another incompatibility, either 
with a non-$*$, or with a 0 or a 1 on a $*$. 
\item[(LPI2)] Pick a $0_i$. 
Then $1_i$ must be incompatible with a non-$*$, or a 0 or a 1 on a $*$.  
\end{itemize}
The first two conditions make $a$ an implicant, and the last two an LPI.
Note also that these conditions are independent. 

Let now $X$ be the number of LPI of length $rn$. We will identify a 
function $L_4=L_4(r,c)$ such that
$$
\frac 1n\log E(X)\le L_4. 
$$
Let 
$$
L_1=L_1(\be,p,z)=z(\be\log p+(1-\be)\log(1-p)-\be\log\be-(1-\be)\log(1-\be)).
$$
This is the exponential rate for the probability that in $zn$ 
Bernoulli trials with success probability $p$ there are exactly $\be n$ 
successes, i.e., this probability is $\approx \exp(L_1n)$. Further, 
if $\kappa, \e,\de\in(0,1)$ are fixed, then among sub-cubes 
with $rn$ non-$*$'s and $\al n$ 1's ($\al\le r$), the proportion 
which have $\e n$ 1's in $[\kappa n, n]$ and $\de n$ $*$'s in 
$[1,\kappa n]$ has exponential rate
$$
\begin{aligned}
L_2=&L_2(r,c,\kappa, \al, \e, \de)\\
=&L_1((\al-\e)/\kappa, \al, \kappa)+L_1(\e/(1-\kappa), \al, 1-\kappa)\\
&+L_1(\de/(\kappa-\al+\e), 1-r, \kappa-\al+\e)+ L_1((1-r-\de)/(1-\kappa-\e), 1-r, 1-\kappa-\e).
\end{aligned}
$$
(Here all four first arguments in $L_1$ are in $[0, 1]$,  
or else the rate is $-\infty$.) 

The expected number of LPI, with $r,\kappa, \e,\de$  given as above, has exponential rate
at most (and this is only an upper bound) 
$$
\begin{aligned}
L_3=&L_3(r,c,\kappa, \al, \e, \de)\\
=&-(1-r)\log(1-r)-\al\log\al-(r-\al)\log(r-\al)\\
&-c(1-r/2)^2\\
&+(r-\al)\log(1-\exp(-c(1-r/2)))\\
&+(\al-\e)\log(1-\exp(-c/2))+\e\log(1-\exp(-c/2)-{\textstyle\frac 12}\de c\exp(-c(1-r/2)))\\
&+L_2(r,c,\kappa, \al, \e, \de). 
\end{aligned}
$$
The next to last line is obtained from (LPI1), as $\e n$ 1's must have 
$\de n$ $*$'s on their left. 

It follows that $L_4$ can be obtained by 
$$
L_4(r,c)=\inf_\kappa\sup_{\al, \e,\de} L_3(r,c,\kappa, \al, \e, \de). 
$$
If $L_4(r,c)<0$, all LPI (for this $c$) \aas~have length at most $r$. Numerical computations
show that this gives a better 
bound than $1-e^{-2c}-ce^{-4c}$ for $c\ge 0.38$. Let us denote the 
best upper bound from the two estimates by $r_u(c)$. This function 
is computed numerically and plotted in Figure 3.  

\begin{figure*}[t]
  \begin{center}
   {\includegraphics[clip, height=5cm]{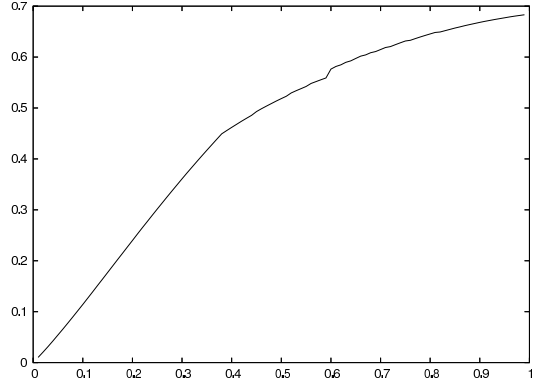}
    }
  \end{center}
  
  \caption{The upper bound $r_u(c)$ for the number of non-$*$'s 
  in the implicant of smallest length included in every cluster 
  of viable genotypes, plotted against $c$.  
  } 
\label{fig_ap_a}
\end{figure*}

\subsection*{Appendix C. Number of clusters under random pair incompatibilities}

In this section we briefly explain why the number of clusters  
under random pair incompatibilities is asymptotically
a function of a Poisson random variable. There is a
clear way to separate the genotype space into disconnected clusters. 
For example, if $F_1=\{(0_1,0_2), (1_2,0_3),(1_1,1_2)\}$, we 
see that every viable genotype has one of these two allele configurations 
on the first two loci: $C=0_11_2$ or $\overline{C}=1_10_2$. 
Since there are no genotypes with $0_10_1$ or $1_11_2$,  
there is no way to mutate from the viable genotypes with $0_11_2$ 
to the viable genotypes with $1_10_2$ without passing through an inviable genotype. 
However, if we add one incompatibility to $F_1$ to make 
$F_2=F_1\cup\{(0_1,1_2)\}$, 
then there are no longer any genotypes with the alleles $0_11_2$
and we return to a single cluster of viable genotypes. 

Notice that the digraph $D_{F_1}$ contains the directed 
cycle $1_1 \to 0_2 \to 1_1$ and equivalently the directed cycle 
$1_2 \to 0_1 \to 1_2$. $D_{F_3}$ also contains these 
cycles but there are paths between them as well: $0_2 \to 0_1$ and $1_1 \to 1_2$.

Formally, a pair of complementary allele configurations 
$(C,\overline{C})$ on a set of  $k \geq 2$ loci is defined to 
be a {\it splitting pair\/} if the digraph $D_F$ contains a directed cycle 
(in any order) on the alleles in $C$ (and equivalently on those in $\overline{C}$,
which consist of reversed alleles in $C$) 
and does not contain a path between the alleles in $C$ and the alleles in $\overline{C}$.
It should be clear from the example $F_1$ above that the existence 
of a splitting pair will create a barrier in the genotype 
space through which it is not possible to pass by mutations on viable genotypes. 
In fact, it is proved in Pitman (unpub.) that 
any two viable genotypes $u$ and $v$ will be disconnected 
in the fitness landscape if and only if the loci on which they 
differ contain a splitting pair.

Thus, the existence of viable genotypes on either side of 
a splitting pair (with each configuration of complementary alleles) 
ensures disconnected clusters. If there are $k$ splitting pairs in the 
formula $F$ and there are viable genotypes with each of the allele 
configurations in each of the splitting pairs then there are $2^k$ clusters 
of viable genotypes. 
The restriction that there be viable genotypes on either side is asymptotically 
unlikely to make a difference as we can   
fix one of the $2^k$ configurations of alleles and \aas~find a   
viable genotype on the remaining loci. Therefore the number of 
clusters of viable genotypes is \aas~equal to $2^X$, where 
$X$ is the number of splitting pairs, provided that $X$ is 
stochastically bounded, but we will see shortly that the expectation
$E(X)$ is bounded. In fact, the next paragraph suggests
that $X$ converges to a  Poisson limiting distribution.
(A detailed discussion of this issue will appear in Pitman (unpub.).)

It follows from \cite{Pal} or \cite{Bol} 
that the number of directed cycles of length $k$ in $D_F$ is 
Poisson$(\lambda_k)$ with $\lambda_k = (2k)^{-1}c^k$. 
In particular, the expected number of splitting pairs converges to
is $\lambda=-\frac{1}{2} (\ln(1-c)+c)$. 
Moreover, the probability that there is no splitting pair   
converges to the product of the probabilities that the cycle of each length is absent
\citep{Pal}, which is
\begin{equation} 
\prod_{k=2}^\infty \exp{\left(-\frac{c^k}{2k}\right)} =
\exp{\left(\frac{ \ln{(1-c)}+c}{2}\right)} = [(1-c)e^c]^{\frac{1}{2}}.
\end{equation} 
In particular, this gives the limiting probability of a unique cluster. 

\subsection*{Appendix D. Proof of equation~(\ref{gamma}).}

In this section we assume that genotypes have multiallelic loci, which are
subject to random pair incompatibilities. The model introduced in Section 5.2
is the most natural, but is not best suited for our second moment approach.
Instead, we will work with the equivalent modified
model with $m$ pair incompatibilities, each 
chosen independently at random, and the first and the second member of each pair  
chosen independently from the $an$ available alleles. We will assume 
that $m=\frac 14ca^2n$, label $c'=\frac 14c$, and denote, as usual, the resulting set 
of incompatibilities by $F$. 

To see that these two models are equivalent for our purposes, 
first note that the number of incompatibilities which are 
{\it not legitimate\/}, in the sense that the two alleles are chosen 
from the same locus, is  stochastically bounded in $n$. (In fact, it 
converges in distribution to a Poisson($c'a^2$) random variable.)
Moreover, by the Poisson approximation to the birthday problem 
\citep{BHJ}, the number of pairs of 
choices which result in the same incompatibility in this model is 
asymptotically Poisson($c'a^2/2$). 
In short, then, the procedure results in the number $m-\cO(1)$ of different legitimate 
incompatibilities. If $m$ in the modified model is increased to, say, $m'=m+n^{2/3}$, then the 
two models could be coupled so that 
the incompatibilities in the original model are included in those in the modified model. As 
the existence of a viable phenotype becomes less likely when $m$ is increased, this demonstrates 
that~(\ref{gamma}) will follow once we show
the following for the modified model: 
for every $\e>0$ there exists a large enough $a$ so 
that $c'<\log a-\e$ implies that  
$N\ge 1$ \aas  

To show this, we introduce the auxiliary random variable 
$$
X=\sum_{\sigma \in \cG_a}\prod_{I\in F}\left(w_01_{\{|I\cap\sigma|=0\}}+
                                        w_11_{\{|I\cap\sigma|=1\}}\right),
$$
where $1_A$ is the indicator of the set $A$.  
The size of the intersection $I\cap\sigma$ is computed by transforming 
both the incompatibility $I$ and 
the genotype $\sigma$ to 
sets of (indexed) alleles, and 
the weights $w_0$ and $w_1$ will be chosen later. To intuitively understand the 
statistic $X$, note that when $w_0=w_1=1$, the product is exactly the indicator of the 
event that $\sigma$ is viable and $X$ is then the number of viable genotypes $N$. In general, 
$X$ gives different scores to different viable genotypes --- however, the crucial fact to note 
is that that $X>0$ iff $N>0$. Therefore 
$$
P(N>0)= P(X>0)\ge (E(X))^2/E(X^2), 
$$
which is how the second moment method is used \citep{AM}. 

As 
$$
\begin{aligned}
&P(|\sigma\cap I|=0)=\left(\frac {a-1}a\right)^2, \\
&P(|\sigma\cap I|=1)=\frac {2(a-1)}{a^2}, \\
\end{aligned}
$$
we have 
$$
E(X)=a^n\left(w_0\left(\frac {a-1}a\right)^2+w_1\frac {2(a-1)}{a^2}\right)^m.
$$
Moreover 
$$
E(X^2)=\sum_{k=0}^n a^n \binom{n}{k}(a-1)^k(w_0^2 P(00)+2w_0w_1P(01)+w_1^2P(11)), 
$$
where $P(01)$ is the probability that $I$ has 
intersection of size $0$ with $\sigma=0_1\dots0_k0_{k+1}\dots 0_n$ and of size $1$ with 
$\tau=1_1\dots1_k0_{k+1}\dots 0_n$, and $P(00)$ and $P(11)$ are defined analogously. Thus, if $k=\al n$, 
$$
\begin{aligned}
&P(00)=\left(1-\frac{1+\al}a\right)^2,\\
&P(01)=\frac{2\al}a\left(1-\frac{1+\al}a\right),\\
&P(11)=\frac{2(1-\al)}a\left(1-\frac{1+\al}a\right)+2\left(\frac\al a\right)^2.
\end{aligned}
$$
Let $\Lambda=\Lambda_{a, w_0, w_1}(\al)$ be the $n$'th root of the 
$k=(\al n)$'th term in the sum for $E(X^2)$, divided by $E(X)^2$. Hence 
$$
\begin{aligned}
\Lambda=&\frac{(a-1)^\al}{a\cdot \al^\al(1-\al)^{1-\al}}\\
&\times \frac 
{\left( w_0^2\left(1-\frac{1+\al}a\right)^2+4w_0w_1\frac{\al}a\left(1-\frac{1+\al}a\right)
        +2w_1^2\left(\frac{(1-\al)}a\left(1-\frac{1+\al}a\right)+\left(\frac\al a\right)^2\right)
 \right)^{c'a^2}}
{\left(w_0\left(\frac {a-1}a\right)^2+w_1\frac {2(a-1)}{a^2}\right)^{2c'a^2}}.
\end{aligned}
$$
Let $\al^*=(a-1)/a$. A short computation shows that $\Lambda=1$ when $\al=\al^*$. 

If $\Lambda>1$ for some $\al$, then $E(X^2)/(E(X))^2$ increases exponentially and 
the method fails (as we will see below, 
this always happens when $w_0=w_1=1$, i.e., 
when $X=N$). On the other hand, if $\Lambda<1$ for $\al\ne\al^*$, and
$\frac{d^2\Lambda}{d\al^2}(\al^*)<0$, then 
Lemma 3 from \cite{AM} implies that $E(X^2)/(E(X))^2\le C$ for some constant 
$C$, which in turn implies that $P(N>0)\ge 1/C$. The sharp threshold 
result then finishes off the proof of~(\ref{gamma}). 

Our aim then is to show that $w_0$ and $w_1$ can be chosen so that, for $c'=\log a-\e$, 
$\Lambda$ has the properties described in the above paragraph.  
We have thus reduced the proof of~(\ref{gamma}) to a calculus problem. 

Certainly the necessary condition is that $\frac{d\Lambda}{d\al}(\al^*)=0$, and
$$
\frac{d\Lambda}{d\al}(\al^*)=-\frac 2{a^3}(w_0(a-1)-w_1(a-2))^2, 
$$
so we choose $w_0=a-2$ and $w_1=a-1$. (Only the quotient between $w_0$ and $w_1$ 
matters, so a single equation is enough.) This simplifies $\Lambda$ to
$$
\Lambda=\Lambda_a(\al)=\frac{(a-1)^\al}{a\al^\al(1-\al)^{1-\al}}
\cdot 
\frac 
{\left(\left(\al -\frac{a-1}a\right)^2-\frac{(a-1)^4}{a^2}\right)^{c'a^2}}
{\left(\frac{(a-1)^2}a\right)^{2c'a^2}}. 
$$
Let $\varphi=\log\Lambda$. We need to demonstrate that $\varphi<0$ for $\al\in[0,\al^*)\cup (\al^*, 1]$ 
and that $\varphi''(\al^*)<0$. A further simplification can be obtained 
by using $x-Cx^2\le \log(1+x)\le x$ (valid for all nonnegative $x$), 
which enables us to transform $\varphi$ (without changing the 
notation) to 
$$
\varphi(\al)=c'\frac{a^4}{(a-1)^4}\left(\al -\frac{a-1}a\right)^2 
-\al \log\al-(1-\al)\log(1-\al)+\al\log(a-1)-\log a.
$$
Now 
$$
\varphi''(\al)=2c'\frac{a^4}{(a-1)^4}-\frac 1{\al(1-\al)}.
$$
So automatically, for $c'$ large but $c'=o(a)$, $\varphi''(\al^*)<0$ for large $a$. Moreover, 
$\varphi$ cannot have another local maximum when $\varphi''>0$. If 
$\varphi(\al)\ge 0$ for some $\al\ne\al^*$, then this must happen for an $\al$ 
in one of the two intervals 
$[0, 1/(2c')+\cO((c')^{-2})]$ or $[1- 1/(2c')-\cO((c')^{-2}), 1]$. 
Now, $\varphi$ has a unique
maximum at $\al^*$ in the second interval. In the first interval, 
a short computation shows that 
$$
\varphi(\al)=-\e-\al \log a+\cO\left(\frac{\log\log a}{\log a}\right), 
$$
which is negative for large $a$. This ends the proof. 

This method yields nontrivial lower bounds for $\gamma$ for all $a\ge 3$, 
cf.~Table 1. 

\begin{center}
\renewcommand{\arraystretch}{.75}
\begin{table}[t]
\caption{The lower bounds on $\gamma$ obtained by the method described in 
text, compared to the easy upper bounds $4\log a$.
}%
\label{t2x1_prel}%
{\normalsize \vspace{.2in} }
\par
\begin{center}
{\normalsize
\begin{tabular} {|r||r|r|}\hline
$a$ & l.~b.~on $\gamma$ & $4\log a$\\ \hline\hline
3 & 1.679 & 4.395\\ 
4 & 2.841 & 5.546\\
5 & 3.848 & 6.438\\
6 & 4.714 & 7.168\\
7 & 5.467 & 7.784\\
8 & 6.128 & 8.318\\
9 & 6.715 & 8.789\\
10 & 7.242 & 9.211\\
20 & 10.672 & 11.983\\
30 & 12.608 & 13.605\\
40 & 13.944 & 14.756\\
50 & 14.960 & 15.649\\
100 & 18.017 & 18.421\\
200 & 20.982 & 21.194\\
300 & 22.663 & 22.816\\
400 & 23.846 & 23.966\\
500 & 24.759 & 24.859\\
\hline
\end{tabular}
}
\end{center}
\end{table}
\end{center}

\subsection*{Appendix E. Existence of viable phenotypes.}

In this section we describe a comparison between models from Sections 5.2 and 5.3
that will yield the result in Section 5.3. 
We begin by assuming that $a=1/r$ is an integer, which we can do without loss of generality.  
Divide the $i$'th coordinate interval $[0,1]$ into $a$ disjoint intervals $I_{i0},\dots, I_{i,{a-1}}$
of length $r$. For a phenotype $x\in \cP$ let $\Delta(x)\in \cG_a$ be determined
so that $\Delta(x)_i=j$ iff $x_i\in I_{ij}$. 

Note that, as soon as $I_{i_1j_1}\times I_{i_2j_2}$ contains a point in 
$\cP_{i_1i_2}$, no $x$ with $\Delta(x)_{i_1}=j_1$ and $\Delta(x)_{i_2}=j_2$
is viable. This happens independently for each such Cartesian product, 
with probability $1-\exp(-\la r^2)\ge cr^2/(2n)$. 
Therefore, using the result from Section 5.2, when $cr^2>4\log a=-4\log r$, there is
\aas~no viable  
genotype. 

On the other hand, let $I^\e$ be the closed $\e$-neighborhood of the interval 
$I$ in $[0,1]$ (the set of points within $\e$ of $I$), and consider
the events that $I_{i_1j_1}^{r/2}\times I_{i_2j_2}^{r/2}$ contains a point in 
$\Pi_{i_1i_2}$. These events are independent if we restrict $j_1,j_2$ 
to even integers. Moreover, each has probability 
$1-\exp(-4\la r^2)\sim 4cr^2/(2n)$, for large $n$. 
It again follows from Section 6.2 that a viable genotype $x$ 
with $\Delta(x)_i$ even for all $i$, \aas~exists as soon as
$4cr^2<4(\log (a/2)-o(1))=(-4\log r-\log 2-o(1))$. 

\end{document}